\documentclass[journal]{IEEEtran}
\ifCLASSINFOpdf
\else
\fi

\usepackage{xfrac}

\usepackage{array}

\usepackage{multirow}

\usepackage{tikz}
\usetikzlibrary{arrows,positioning} 

\usepackage{amsmath}
\usepackage{amsthm}
\theoremstyle{definition}

\newtheorem{definition}{Definition}

\usepackage{algpseudocode}
\usepackage{algorithm}

\usepackage[caption=false]{subfig}

\def\schemNameUnformatted{FogPlan}
\def\schemeName{{\sc\schemNameUnformatted}}
\def\qdfsp{QDFSP}
\def\qdfspFull{QoS-aware Dynamic Fog Service Provisioning (\qdfsp)}
\def\myjournaltitle{\schemeName: A Lightweight QoS-aware Dynamic Fog Service Provisioning Framework}

\usepackage[pscoord]{eso-pic}
\newcommand{\placetextbox}[3]{
\setbox0=\hbox{#3}
\AddToShipoutPictureFG*{ \put(\LenToUnit{#1\paperwidth},\LenToUnit{#2\paperheight}){\vtop{{\null}\makebox[0pt][c]{#3}}}
}
}
\placetextbox{.5}{0.055}{\scriptsize{Copyright (c) 2018 IEEE. Personal use is permitted. For any other purposes, permission must be obtained from the IEEE by sending a request to pubs-permissions@ieee.org.}}

\begin{document}
%
\title{\myjournaltitle}

\author{Ashkan~Yousefpour,~\IEEEmembership{Graduate Student Member,~IEEE,} Ashish~Patil, Genya Ishigaki,~\IEEEmembership{Graduate Student Member,~IEEE,} Inwoong~Kim,~\IEEEmembership{Member,~IEEE,} Xi~Wang,~\IEEEmembership{Member,~IEEE,} Hakki~C.~Cankaya,~\IEEEmembership{Member,~IEEE,} Qiong~Zhang,~\IEEEmembership{Member,~IEEE,} Weisheng Xie,~\IEEEmembership{Member,~IEEE,} Jason~P.~Jue,~\IEEEmembership{Senior Member,~IEEE,} 
\thanks{Manuscript received October 26, 2018; revised January 9, 2019; accepted January 24, 2019. Date of publication Month ??, 2018; date of current version Month ??, 2018.  ({\em Corresponding author: Ashkan~Yousefpour.})}
\thanks{Ashkan~Yousefpour, Ashish~Patil, Genya Ishigaki, and Jason~P.~Jue are with the Department of Computer Science, The University of Texas at Dallas, Richardson, TX, 75080 USA. 
(e-mail: ashkan@utdallas.edu; axp175330@utdallas.edu; gishigaki@utdallas.edu; jjue@utdallas.edu)}

\thanks{Inwoong~Kim, Xi~Wang, and Qiong~Zhang are with Fujitsu Laboratories of America, Richardson, TX 75082 USA, (e-mail: inwoong.kim@us.fujitsu.com; xi.wang@us.fujitsu.com; qiong.zhang@us.fujitsu.com)}

\thanks{Hakki~C.~Cankaya and Weisheng Xie are with Fujitsu Network Communications, Richardson, TX 75082 USA, (e-mail: hakki.cankaya@us.fujitsu.com; wilson.xie@us.fujitsu.com)}

\thanks{Digital Object Identifier ??.????/IOTJ.2019.???????}
}


%


\newenvironment{editted}{\color{blue}}{}

\maketitle

\begin{abstract}
Recent advances in the areas of Internet of Things (IoT), Big Data, and Machine Learning have contributed to the rise of a growing number of complex applications. These applications will be data-intensive, delay-sensitive, and real-time as smart devices prevail more in our daily life. Ensuring Quality of Service (QoS) for delay-sensitive applications is a must, and fog computing is seen as one of the primary enablers for satisfying such tight QoS requirements, as it puts compute, storage, and networking resources closer to the user.


In this paper, we first introduce \schemeName, a framework for \qdfspFull. \qdfsp~concerns the dynamic deployment of application services on fog nodes, or the release of application services that have previously been deployed on fog nodes, in order to meet low latency and QoS requirements of applications while minimizing cost. \schemeName~framework is practical and operates with no assumptions and minimal information about IoT nodes. Next, we present a possible formulation (as an optimization problem) and two efficient greedy algorithms for addressing the \qdfsp~at one instance of time. Finally, the \schemeName~framework is evaluated using a simulation based on real-world traffic traces.
\end{abstract}


\section{Introduction}
The Internet of Things (IoT) is shaping the future of connectivity, processing, and reachability. In IoT, every ``thing'' is connected to the Internet, including sensors, mobile phones, cameras, computing devices, and actuators. IoT is proliferating exponentially, and IoT devices are expected to generate massive amounts of data in a short time, with such data potentially requiring immediate processing. 

Many IoT applications, such as augmented reality, connected and autonomous cars, drones, industrial robotics, surveillance, and real-time manufacturing have strict latency requirements, in some cases below 10 ms \cite{byers2017architectural}\cite{extendingCloudtoEdge}. These applications cannot tolerate the large and unpredictable latency of the cloud when cloud resources are deployed far from where the application data is generated. Fog computing \cite{fog-survey}, edge computing \cite{edge-survey}, and MEC \cite{taleb2017multi} have been recently proposed to bring low latency and reduced bandwidth to IoT networks, by locating the compute, storage, and networking resources closer to the users. Fog can decrease latency, bandwidth usage, and costs, provide contextual location awareness, and enhance QoS for delay-sensitive applications, especially in emerging areas such as IoT, Internet of Energy, Smart City, Industry 4.0, and big data streaming \cite{openfog}\cite{NIST}.

Certain IoT applications are bursty in resource usage, both in space and time dimensions. For instance, in {\em situation awareness} applications, the cameras in the vicinity of an accident generate more requests than the cameras in other parts of the highway (space), while security motion sensors generate more traffic when there is suspicious activity in the area (time). The resource usage of the delay-sensitive fog applications may be dynamic in time and space, similar to that of situational awareness applications \cite{saurez2016incremental}. 

An IoT application may be composed of several {\em services} that are essentially the components of the application and can run in different locations. Such services are normally implemented as containers, virtual machines (VMs), or unikernels. For instance, an application may have services such as authentication, firewall, caching, and encryption. Some application services are delay-sensitive and have tight delay thresholds, and may need to run closer to the users/data sources (e.g. caching), for instance at the edge of the network or on fog computing devices. On the other hand, some application services are delay-tolerant and have high availability requirements, and may be deployed farther from the users/data sources along the fog-to-cloud continuum or in the cloud (e.g. authentication). We label these fog-capable IoT services that could run on the fog computing devices or the cloud servers as {\em fog services}.

Placement of such services on either fog computing devices ({\em fog nodes}) or cloud servers has a significant impact on network utilization and end-to-end delay. Although cloud computing is seen as the dominant solution for dynamic resource usage patterns, it is not able to satisfy the ultra-low latency requirements of specific IoT applications, provide location-aware services, or scale to the magnitude of the data that IoT applications produce \cite{extendingCloudtoEdge}\cite{reviewer-asked-us-to-cite-his-work}. This is also evident by the recent edge computing frameworks (Microsoft Azure IoT Edge \cite{azure}, Google Cloud IoT Edge \cite{gc}, and Amazon AWS Greengrass \cite{aws}) introduced by major cloud computing companies to bring their cloud services closer to the edge, on the IoT devices.

Fog services could be deployed on the fog nodes to provide low and predictable latency, or in the cloud to provide high availability with lower deployment cost. The placement of the fog services could be accomplished in a static manner (e.g. with the goal of minimizing the total cost while satisfying the corresponding latency constraints). Nevertheless, since the resource usage pattern of certain IoT application is dynamic and changing over time, a static placement would not be able to adapt to such changes. Thus, fog services should be placed dynamically in order to address the bursty resource usage patterns of the overlaying fog-based IoT applications. This is the essence of dynamic placement of fog services, which is to dynamically deploy and release IoT services on fog computing devices or cloud servers to minimize the resource cost and meet the latency and QoS constraints of the IoT applications.

In this paper, we introduce \myjournaltitle. In the next section, we discuss the related research studies and discuss how the \qdfsp~problem fits in the IoT-fog-cloud architecture. We then present a possible formulation of an Integer Nonlinear Programming (INLP) task to address the \qdfsp~problem at one time instance (Section \ref{QDFSP-INLP}), propose two practical greedy algorithms for dynamically adjusting service placement (Section \ref{Heuristic}), and evaluate the algorithms numerically\footnote{The QoS in this paper is modeled in terms of latency, namely the IoT service delay and latency threshold.} (Section \ref{results}). We present conclusions and suggestions for future research in Section \ref{Conclusion}.

\section{Related Work} \label{related}
Some frameworks are conceptually similar to \schemeName~but with goals that differ from the goal of meeting the ultra-low latency requirements of IoT applications, that is the goal of \schemeName. Service migration in edge clouds in response to user movement \cite{abdelwahab2016flocking}, network workload performance \cite{zhang2016segue}, and for reducing file system transfer size \cite{ma2018efficient}; and VM migration and handoff in edge clouds \cite{clark2005live}\cite{thesis-migration}\cite{ha2017you}\cite{service-migration}\cite{chaufournier2017fast} are the most notable among these frameworks. Comparably, the studies \cite{oci}\cite{platform}\cite{Indiefog}\cite{wang2017enorm}\cite{incremental} propose deployment platforms and programming models for service provisioning in the fog. Similarly, a framework and software implementation for dynamic service deployment based on availability and processing resources of edge clouds are presented in \cite{picasso}. To model resource cost in edge networks for fog service provisioning, the authors in \cite{zenith} propose a model for resource contract establishment between edge infrastructure provider and cloud service providers based on auctioning.

Some studies proposed frameworks for fog service provisioning for specific use cases and applications. The authors in \cite{zhang2017towards} propose an edge cloud architecture for gaming that places local view change updates and frame rendering on edge clouds, and global game state updates on the central cloud. They further propose a service placement algorithm for multiplayer online games that periodically makes placement decisions, based on QoS, mutual impact among players, and player mobility patterns. MigCEP \cite{migcep} is proposed for placement and migration of operators in real-time event processing applications over the fog and the cloud. A real-time event processing application is modeled as a set of operators and MigCEP places the operators based on the mobility of users.

Closely related to the \schemeName~framework are schemes for dynamic resource reconfiguration in fog computing. The authors in \cite{dynamic-module-deploy} implement a fog computing platform that dynamically pushes software modules to the fog nodes, whereas the study \cite{asymptotically} provides theoretical foundations and algorithms for storage- and cost-aware dynamic service provisioning on edge clouds. The proposed frameworks, however, are oblivious to QoS and latency, as opposed to the proposed \schemeName~framework. 

Another related and similar framework to \schemeName~is service placement in distributed and micro clouds \cite{placement1}\cite{placement2}\cite{placement3} and virtual network embedding \cite{vne1}. Nevertheless, ultra-low delay requirements and delay violation metrics are not considered in these studies, as opposed to \schemeName. The study in \cite{skarlat2017towards} proposes a framework for the fog service provisioning problem. The proposed framework, however, does not consider the overhead of the introduced components on IoT nodes and assumes that IoT nodes run fog software and a virtualization technology. Similar to fog service provisioning, the work in \cite{yigitoglu2017foggy} introduces IoT application provisioning by proposing a service-oriented mobile platform.  
\begin{figure}[!t]
\centering
    \includegraphics[width=1\linewidth]{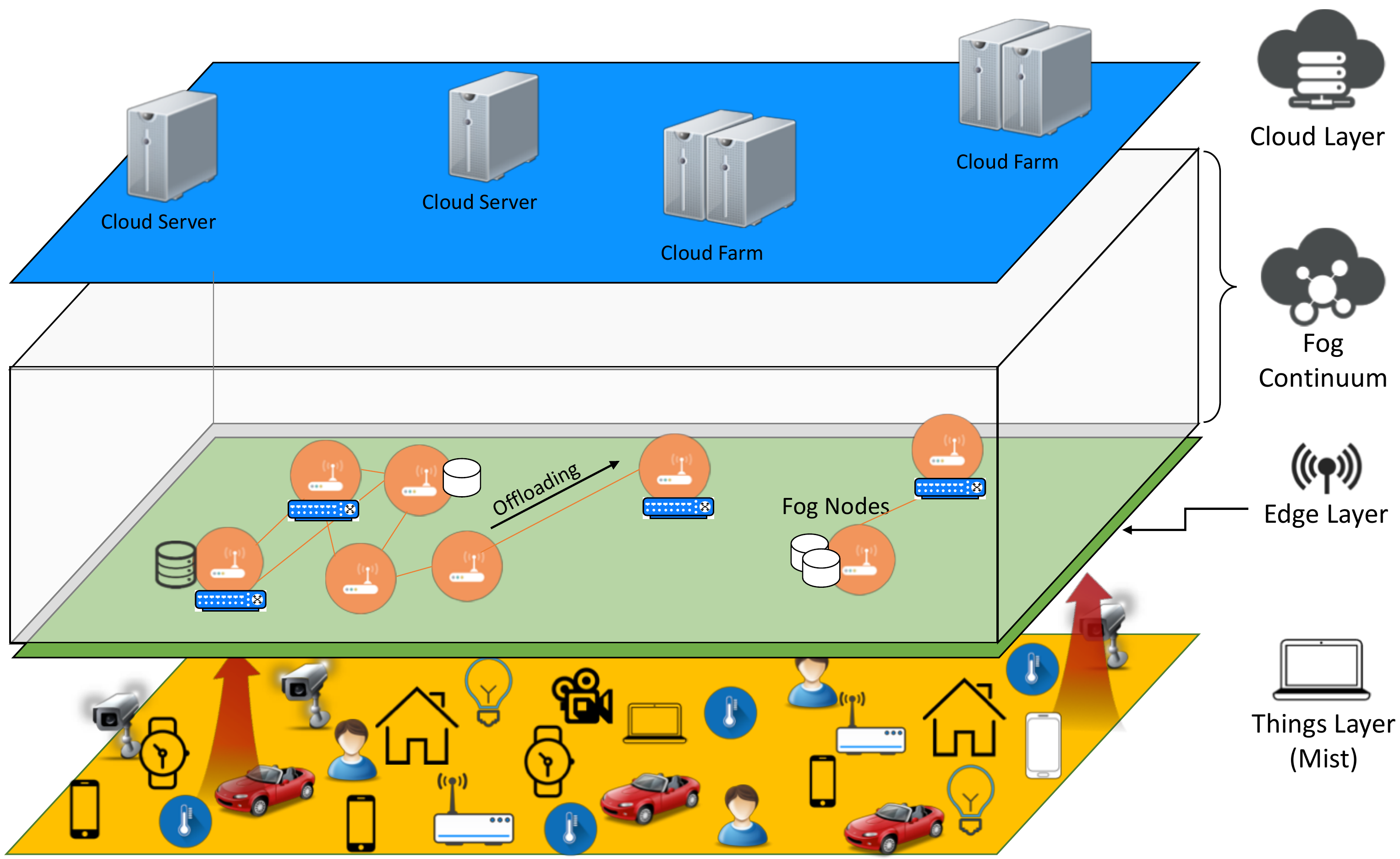}
    \caption{General architecture for IoT-fog-cloud. The bottom logical layer (yellow) is the things layer, where IoT devices lie. The top logical layer (blue) is the cloud layer. Fog is the continuum between things and the cloud layer, and it is not only limited to the edge. (Edge computing occurs at the edge layer (green), while fog computing can occur anywhere along the IoT to cloud). 
}
    \label{fig:architecture}
\end{figure}
Despite the substantial contributions in the studies mentioned, they cannot be directly applied to the dynamic provisioning of delay-sensitive fog-capable IoT applications. Firstly, \schemeName~is not application-specific and is designed for general fog-capable IoT applications. Secondly, it does not require any knowledge of user mobility patterns or status/specifications of the IoT devices. Moreover, \schemeName~does not make any assumptions about IoT devices and their supported software capabilities (e.g., virtualization). The \schemeName~framework is QoS-aware and considers ultra-low delay requirements of IoT applications and customer delay violation metrics. Finally, \schemeName~is lightweight and designed to scale to the large magnitude of IoT networks. Each of the discussed studies considers some of the above features, whereas \schemeName~considers all of them collectively, to fill the gap in the literature for a lightweight framework for dynamic provisioning of delay-sensitive fog-capable IoT applications.  

\subsection{Summary of Contributions}
The contributions of this paper include: (1) \schemeName, a novel, lightweight, and practical framework for QoS-aware dynamic provisioning (deploying and releasing) of fog services with no assumptions and minimal information about IoT nodes, (2) a formulation of an optimization problem for the \qdfsp~problem at one time instance, (3) two efficient greedy algorithms (one with regards to on QoS and one with regards to cost) are proposed for addressing the \qdfsp~problem periodically, and (4) evaluation based on real-world traffic traces to verify the applicability and scalability of the \schemeName. 

\section{\schemeName~Framework} \label{model}
The main goal of \schemeName~is to provide better QoS, in terms of reduced IoT service delay and reduced fog-to-cloud bandwidth usage, while minimizing the resource usage of the Fog Service Provider (FSP). Specifically, the QoS improvement is useful for low latency and bandwidth-hungry applications such as real-time data analytics, augmented and virtual reality, fog-based machine learning, ultra-low latency industrial control, or big data streaming. 
\schemeName~is a framework for the \qdfsp~problem, which is to dynamically provision (deploy or release) IoT services on the fog nodes or cloud servers, in order to comply with the QoS terms (delay threshold, penalties, etc.) defined as an agreement between the FSP and its clients, with minimal resource cost for the FSP.  

\subsection{IoT-Fog-Cloud Architecture}
In this subsection, we discuss the general architecture for an IoT-fog-cloud system, define {\em fog nodes}, show how requests from IoT devices are handled, and describe the available communication technologies between nodes in the general IoT-fog-cloud architecture.
\subsubsection{Layered Architecture}
Fig. \ref{fig:architecture} shows the general architecture of IoT-fog-cloud. At the bottom is the {\em things layer}, where IoT devices logically lie. The things layer is sometimes referred to as the {\em mist layer} since it is where mist computing occurs. Mist computing captures a computing paradigm that occurs on the connected devices \cite{mist-cisco}. IoT devices are endpoint devices such as sensors, actuators, mobile phones, smart fridges, cameras, smart watches, cameras, and autonomous cars. The next logical layer is the {\em edge layer}, where {\em edge devices} such as WiFi access points, first hop routers and switches, and base stations are located. Edge computing normally takes place on the (edge) devices of this layer. 

At the top is the {\em cloud layer}, where large-scale cloud servers and well-provisioned data centers are located. The cloud servers and data centers in the cloud layer are normally far from the IoT devices. The {\em fog continuum} (where fog computing occurs) fills the computing, storage, and decision making the gap between the things layer and the cloud layer \cite{openfog}. The fog continuum is not only limited to the edge layer as it includes the edge layer and expands to the cloud. Fog computing is hierarchical and it provides computing, networking, storage, and control anywhere from the cloud layer to the things layer; while, edge computing tends to be the computing at the edge layer \cite{openfog}\cite{NIST}. 
\subsubsection{Fog Node}
We refer to the devices in the fog continuum as {\em fog nodes}. Fog nodes provide networking functions, computation, and storage to the IoT devices in the things layer and are located closer than cloud servers to the IoT devices \cite{NIST}. Fog nodes could host services packaged in the form of VMs, containers, or unikernels. Fog nodes can be routers, switches, dedicated servers for fog computing (e.g. cloudlets), customer premises equipment (CPE) nodes (also called home gateways), or firewalls  \cite{NIST}\cite{NetFATE}\cite{faraci2015analytical}. The fog nodes can be either small-scale nodes in a residential environment (e.g. home gateways), or medium/ high-performance equipment in an enterprise environment (e.g. routers or aggregation nodes of a Telco network) \cite{NetFATE}.

\subsubsection{Handling Requests in IoT-fog-cloud}
IoT nodes send their requests to the cloud. The IoT requests that are for traditional cloud-based services are sent directly to the cloud without any processing in the fog. In this case, the requests are sent to the cloud and may bypass fog nodes along the path, but without any fog processing. On the other hand, the IoT requests that are for fog services will either be processed by fog nodes, if the corresponding service is deployed on the fog nodes, or forwarded to the cloud if the service is not deployed on fog nodes. Fog nodes may also offload requests to other fog nodes, such as in \cite{ashkan-fog-delay}, to balance their workload and to minimize response delay of the IoT requests. 

\subsection{Clients' QoS Requirements}
\subsubsection{QoS Measure}
In this paper, we focus on the average IoT service delay as the QoS measure of IoT nodes; the average service delay is compared against a desired service delay threshold as the QoS requirements. These delay parameters along with few other metrics (such as penalty cost for delay violation) are used to model QoS. The service delay is formulated in section \ref{constraints}.  

\subsubsection{Client}
In this framework, the client is defined as an IoT application developer, IoT application owner, or the entity who owns the IoT nodes and agrees with a level of QoS requirements with the FSP. When a client signs a contract with the FSP, the FSP guarantees to provide fog services (using \schemeName) with service delays below a maximum delay threshold $th$. The FSP also provides a level of the desired QoS $q$, which governs how strict the delay requirements are. For example, for a given service $a$, if $q_a=99\%$ and $th_a=5$ ms, it means that the client requires that the service delay of service $a$ must be less than 5 ms, 99\% of the time. All parameters of the framework are summarized in Table \ref{parameters}.

\subsubsection{Complying with QoS Requirements}
The FSP may own fog and edge resources, or it may rent them from edge network providers (e.g. AT\&T, Nokia, Verizon) or edge resource owners \cite{zenith}\cite{edge-resource}\cite{ERO}. To comply with the QoS terms, one trivial solution for the FSP is to blindly deploy the particular service on all the fog nodes close to the client's IoT nodes. Nevertheless, this is not an efficient solution and it wastes the available resources, because it over-provisions resources for that particular service. If the FSP has many clients, it is likely that it does not have adequate resources on fog nodes for all the clients' services. Therefore, the FSP should employ a dynamic approach (e.g., \schemeName) to provision its services, while minimizing the cost and complying with the QoS requirements of its clients. 

\begin{figure}[!t]
\centering
    \includegraphics[width=0.85\linewidth]{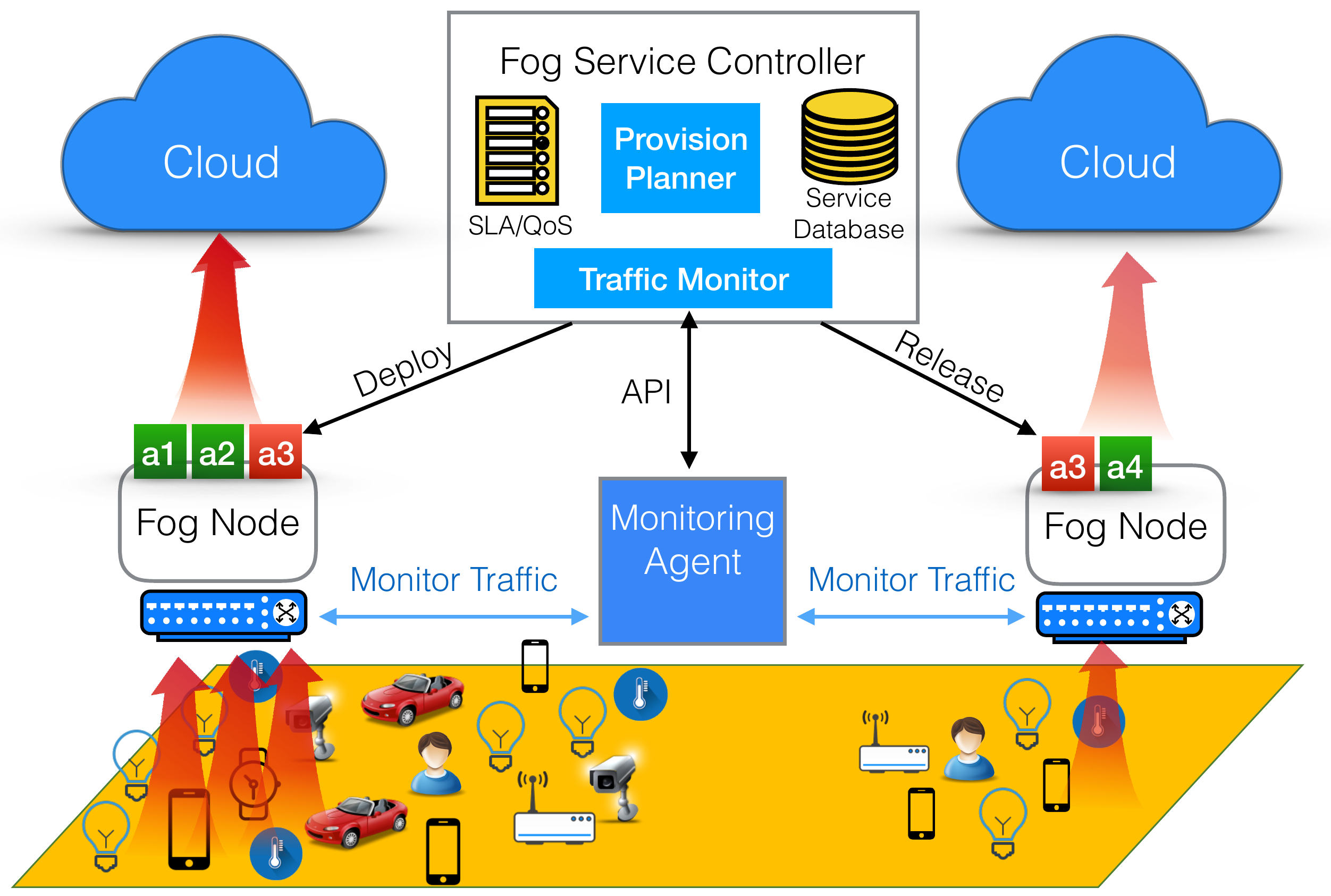}
    \caption{\schemeName~framework. (Only traffic from IoT to cloud is shown.)}
    \label{fig:deploy}
\end{figure}

\subsection{Fog Service Controller (FSC)} \label{fsc}
\subsubsection{General Architecture}
The general architecture of the system for dynamically provisioning the fog services is shown in Fig. \ref{fig:deploy}. The aggregated incoming traffic rate of IoT requests to a fog node is monitored by a traffic monitoring agent, such as that of a Software Defined Networking (SDN) controller. Commercial SDN controllers such as OpenDayLight and ONOS have rich traffic monitoring northbound APIs for monitoring the application-level traffic \cite{sdn-monitor, sdn-monitor2}. The traffic rate monitoring is done in the {\em Traffic Monitor} module. 

The ``{\em Fog Service Controller}'' (FSC) is where \schemeName~is implemented. Essentially, FSC solves the \qdfsp~problem using the monitored incoming traffic rate to the fog nodes and other parameters to make decisions for deploying or releasing services. It is assumed that the FSC only manages fog nodes in a particular geographical location; however, the FSC may be replicated for different geographical areas. 

\subsubsection{What is Inside?}
The FSC has a database (labeled {\em Service Database}) of the application services (e.g. containers) that the FSP's clients develop, and also maintains the QoS parameters of each service (Fig. \ref{fig:deploy}). The FSC uses the Service Database to deploy required services on the fog nodes. This is analogous to the container registry \cite{container-registry} of the recently proposed Microsoft Azure IoT Edge framework. Using QoS parameters, the FSC can obtain the delay and other parameters of a service. When the traffic demand for a particular service increases, the {\em Provision Planner} module in the FSC may deploy a new service on the corresponding fog nodes to reduce the IoT service delay. On the other hand, if demand for a particular service is not significant, the Provision Planner may release the service to save on resource cost. Both the deploy and the release operations are performed as per the QoS requirements. For instance, when the traffic demand is low but the QoS requirement of a service is strict, the service may not be released. 

\subsection{Scalability and Practicality}
The \schemeName~framework is lightweight and practical, and it operates with no assumptions and with minimal information about IoT nodes. The only information from IoT nodes that is required for the purpose of dynamic fog service provisioning is the aggregated incoming traffic rate of IoT nodes to the fog nodes ($\lambda^{\text{in}}_{aj}$). Depending on the definition scope of service delay (to be discussed in Sections \ref{service-delay-equation} and \ref{delay-budget}), the average transmission rate $r_{(I_a,j)}$ and propagation delay $d_{(I_a,j)}$ between the IoT nodes running a service and their corresponding fog node may also be required. The aggregated traffic rate can be easily monitored, while the average transmission rate and the propagation delay can be approximated by round-trip delay, which can be measured using a simple ping mechanism. Nevertheless, if obtaining average values for $r_{(I_a,j)}$ and $d_{(I_a,j)}$ is not possible in a scenario, the definition scope of service delay can be altered (refer to Section \ref{delay-budget}). Knowing minimal information about IoT nodes makes our framework practical and lightweight. Moreover, there are no assumptions about IoT nodes, namely, if they run any specific fog-related application or protocol, support virtualization technology, or have specific hardware characteristics.  

\subsection{Application Development and Deployment}
The communication of the FSC with other entities can be encrypted for added security. Also, adequate secure techniques may be used for monitoring IoT traffic. The security considerations of \schemeName~are outside the scope of this paper and are left as future work. When the FSP's clients develop new services and push them to the cloud, the services are automatically pulled into the FSC's Service Database (encrypted channel). This ensures that the Service Database always has the newest version of the services. This deployment model hides the platform heterogeneity, location of fog nodes, and the complexities of the migration process of the services from the application developers. Note that traditional cloud-based services that do not require the low latency of fog need not be pulled into the Service Database. 

When a service is deployed on a fog node, it advertises the fog node's IP address to the participating IoT nodes that run the application, so that the IoT node's requests are sent to the fog node (fog discovery). Another possibility is for the IoT nodes to discover the new fog service through a service discovery protocol, or similar to the PathRoute module in \cite{Cloudpath}, route the requests to fog services by including a URI in the requests. One can design a fog service discovery protocol as an extension to \schemeName. 

\begin{table}[t]
    \caption{Table of Notation}
    \newcolumntype{L}[1]{>{\raggedright\let\newline\\\arraybackslash\hspace{0pt}}m{#1}}
    \centering
    {\footnotesize
        {\renewcommand\arraystretch{1.0} 
            \begin{tabular}{|c||L{0.8\columnwidth}|}
                \hline
                $F$ & set of fog nodes\\
                $C$ & set of cloud servers\\
                $A$ & set of fog services\\
                $I_a$ & set of IoT nodes running service $a$\\
                $\Phi$ & Fog Service Controller (FSC)\\ 
                \hline
                $q_a$ & desired quality of service for service $a$; $q_a\in (0,1)$ \\
                $th_a$ & delay threshold for service $a$\\
                $d_{aj}$ & average service delay of IoT nodes served by fog node $j$ for service $a$\\
                $p_a$ & penalty that FSP must pay per request of service $a$ if delay requirements are violated by 1\% (in dollar per request per \%)\\
                $\tau_a$ & time interval between two instances of solving optimization problem or greedy algorithms for service $a$ (in seconds)\\
                \hline
                $u_{(s,d)}$ & communication cost of link $(s,d)$ per unit bandwidth per sec\\
                $r_{(s,d)}$ & transmission rate (bandwidth) of link $(s,d)$\\
                $d_{(s,d)}$ & propagation delay of link $(s,d)$\\
                \hline
                $C_j^P$ & unit cost of process. at fog node $j$ (per million instructions)\\
                $C'^P_k$ & unit cost of process. at cloud ser. $k$ (per million instructions)\\
                $C_j^S$ & unit cost of storage at fog node $j$ (per byte per second)\\
                $C'^S_k$ & unit cost of storage at cloud server $k$ (per byte per second)\\
                $L_a^S$ & storage size of service $a$, in bytes\\

                $L_a^P$ & required amount of processing for service $a$ per request (in million instructions per request)\\
                $L_a^M$ & required amount of memory for service $a$ (in bytes)\\
                \hline
                $\lambda^{\text{out}}_{aj}$ & rate of dispatched traffic for service $a$ from fog node $j$ to the associated cloud server (request/second)\\
                $\lambda^{\text{in}}_{aj}$ & incoming traffic rate from IoT nodes to fog node $j$ for service $a$ (request/second)\\
                $\lambda'^{\text{in}}_{ak}$ & incoming traffic rate to cloud server $k$ for service $a$  (request/second)\\
                $\lambda_{ajj'}$ & traffic rate of service $a$ from fog node $j$ to fog node $j'$ (request/second)\\
                \hline
                $h_a(j)$ & index of the cloud server to which the traffic for service $a$ is routed from fog node $j$\\
                $H_a^{-1}(k)$ & set of indices of all fog nodes that route the traffic for service $a$ to cloud server $k$. $H_a^{-1}(k)=\{j|h_a(j)=k\}$\\
                                \hline
                $K^S_j$ & storage capacity of fog node $j$, in bytes\\
                $K^P_j$ & processing capacity (maximum service rate) of fog node $j$, in MIPS\\
                $K^M_j$ & memory capacity of fog node $j$, in bytes\\
                $K'^S_k$ & storage capacity of cloud server $k$, in bytes\\
                $K'^P_k$ & processing capacity (maximum service rate) of cloud server $k$, in MIPS\\
                $K'^M_k$ & memory capacity of cloud server $k$, in bytes\\
                \hline
                $n_j$ & number of processing units of fog node $j$\\
                $n'_k$ & number of processing units of cloud server $k$\\
                $\mu_j$ & service rate of one processing unit of fog node $j$ (in MIPS)\\
                $\mu'_k$ & service rate of one processing unit of cloud server $k$ (in MIPS)\\
                \hline
                $l^{rq}_a$ & average size of requests of service $a$, in bytes\\
                $l^{rp}_a$ & average size of responses of service $a$, in bytes\\
                \hline
                $w_{aj}$ & waiting time (processing delay plus queueing delay) for requests of service $a$ at fog node $j$\\
                $w'_{ak}$ & waiting time (processing delay plus queueing delay) for requests of service $a$ at cloud server $k$\\
                \hline
                $x_{aj}$ & binary variable showing if service $a$ is hosted on fog node $j$\\
                $x'_{ak}$ & binary variable showing if service $a$ is hosted on cloud server $k$\\
                \hline
                $V^\%_a$ & the percentage of IoT service delay samples of service $a$ that do not meet the delay requirement.\\
                \hline
            \end{tabular}
        }}
        \label{parameters}
    \end{table}

\subsection{Fog Node Architecture}
\subsubsection{Monitoring}
Monitoring the incoming traffic to fog nodes is necessary for the operation of \schemeName. Fog nodes may be devices such as switches or routers that already have monitoring capabilities. If fog nodes do not have monitoring capabilities, they must be either directly connected to monitoring-enabled devices or have monitoring software agents running on them. The FSC's Traffic Monitor module will then communicate with the fog nodes (either directly or through the fog nodes' monitoring agent) to get the incoming traffic to the fog nodes. 

\subsubsection{Virtualziation}
Fog nodes in our \schemeName~framework are assumed to run some container orchestration software, such as Docker, Kubernetes, or OpenStack, which automates the deployment and release of service containers. The reason for using containers instead of traditional VMs is that containers are light-weight in comparison to VMs and provide lower set-up delay, as they share the host OS \cite{container-as-service}\cite{dynamic-module-deploy}. We consider {\em stateless} fog services in the form of containers in our framework, so that deploying and releasing them is fast and on-demand, as opposed to slow migration procedures present in VM-based migration techniques \cite{Cloudpath, ha2017you, chaufournier2017fast}. We leave the consideration of stateful fog services and their related migration issues as future work. 
\subsection{The \qdfsp~Problem} \label{QDFSP-INLP}
In this subsection, we formally define the \qdfsp~Problem.
\begin{definition}
    {\em \qdfspFull} is an online task that aims to dynamically provision (deploy or release) services on fog nodes, in order to comply with the QoS requirements while also minimizing the cost of resources. 
\end{definition}
In the next two sections, we first present a possible formulation of an optimization problem for the \qdfsp~problem at one time instance. Then, we present two greedy algorithms that are periodically called to address the \qdfsp~problem.

\section{Addressing \qdfsp~in One Instance of Time as INLP}
In this subsection, we present a possible formulation of the associated optimization problem to address the \qdfsp~periodically, that considers the service provisioning at a given point in time. All notation is explained in Table \ref{parameters}. Let a set of fog nodes be denoted by $F$, a set of cloud servers by $C$, and a set of fog services by $A$. Let the desired QoS level for service $a$ be denoted by $q_a\in (0,1)$, and delay threshold for service $a$ by $th_a$.  

The underlying communication network, a set of fog nodes and cloud servers are modeled as a graph $G=(V,E)$, such that the node set $V$ includes the fog nodes $F$ and cloud servers $C$ ($V=F\cup C$), and the edge set $E$ includes the logical links between the nodes. Fog nodes and cloud servers could be located anywhere, and there is no restriction on the physical network topology. Each edge $e(src,dst)\in E$ is associated with three numbers: $u_e$, the communication cost of logical link $e$ (cost per unit bandwidth used per unit time, that is $\sfrac{\text{cost}}{\text{byte}}$); $r_e$, the transmission rate of logical link $e$ (megabits per second); and $d_e$, the propagation delay of logical link $e$ (milliseconds). These parameters are maintained by the Network Provider (NP) and are shared with the FSP.

The main decision variables of the optimization problem are the placement binary variables, defined below:
\begin{IEEEeqnarray}{rCl} 
     x_{aj}&=&
    \begin{cases}
        1, & $if service $a$ is hosted on fog node $j$$,\\
        0, & $otherwise$.
    \end{cases}\\
    x'_{ak}&=&
    \begin{cases}
        1, & $if service $a$ is hosted on cloud server $k$$,\\
        0, & $otherwise$.
    \end{cases}
 \end{IEEEeqnarray}
Incidentally, we denote by $x^{\text{cur}}_{aj}$, the current placement of the service $a$ on fog node $j$, which can be regarded as an input to the optimization problem to find the future placement of services on fog nodes ($x_{aj}$) and cloud nodes ($x'_{ak}$). 

The binary variable $x'_{ak}$ denotes placement of service $a$ for each cloud server $k$. One can simplify this formulation by removing the index $k$, considering all cloud servers as a whole. To be more general, we consider the first formulation ($x'_{ak}$).

\subsection{Objective}
The optimization problem can be formulated as problem \textbf{P1}:

\begin{IEEEeqnarray}{rCl} 
     \textbf{P1}: &\text{Minimize}& (C^{\text{proc}}_{C}+C^{\text{proc}}_{F})+(C^{\text{stor}}_{C}+C^{\text{stor}}_{F})+\IEEEnonumber\\ 
     &~~&(C^{\text{comm}}_{FC}+C^{\text{comm}}_{FF}+C^{\text{depl}}_{\Phi F})\IEEEnonumber\\ 
     &\text{~Subject~to}&\text{~QoS~constraints}. \IEEEnonumber
 \end{IEEEeqnarray}

The cost components are defined below.

\begin{IEEEeqnarray}{rCl} 
C^{\text{proc}}_{C}&=&\sum_{k\in C}{\sum_{a\in A}{C'^P_k L^P_a\lambda'^{\text{in}}_{ak} x'_{ak}}}\tau_a, \label{cost-proc-cloud}\\
    C^{\text{proc}}_{F}&=&\sum_{j\in F}{\sum_{a\in A}{C^P_j L^P_a \lambda^{\text{in}}_{aj} x_{aj}}}\tau_a,\\
    C^{\text{stor}}_{C}&=&\sum_{k\in C}{\sum_{a\in A}{C'^S_k L^S_a x'_{ak}}}\tau_a,\\
    C^{\text{stor}}_{F}&=&\sum_{j\in F}{\sum_{a\in A}{C^S_j L^S_a x_{aj}}}\tau_a,\\
    C^{\text{comm}}_{FC}&=&\sum_{j\in F}{\sum_{a\in A}{u_{(j,h_a(j))} \lambda^{\text{out}}_{aj}(l^{rq}_a+l^{rp}_a)} \tau_a},\\
    C^{\text{comm}}_{FF}&=&\sum_{j\in F}{\sum_{j'\in F}{\sum_{a\in A}{u_{(j,j')} \lambda_{ajj'}}l^{rq}_a} \tau_a},\\
        C^{\text{depl}}_{\Phi F}&=&\sum_{j\in F}{\sum_{a\in A}{u_{(\Phi,j)}(1-x^{\text{cur}}_{aj})x_{aj}L^S_a }}.
\end{IEEEeqnarray}

$C^{\text{proc}}_{C}$ and $C^{\text{proc}}_{F}$ are cost of processing in cloud and fog, respectively; $C^{\text{stor}}_{C}$ and $C^{\text{stor}}_{F}$ are cost of storage in cloud and fog, respectively. $C^{\text{comm}}_{FC}$ is the cost of communication between fog and cloud, $C^{\text{comm}}_{FF}$ is the cost of communication between fog nodes, and $C^{\text{depl}}_{\Phi F}$ is the communication cost of service deployment, from the FSC $\Phi$ to fog nodes. A service deployed on a fog node may be released when the demand for the service is small. Therefore, we assume the services are stateless, that is they do not store any state information on fog nodes \cite{extendingCloudtoEdge}, and we do not consider costs for state migrations. We consider a discrete-time system model where time is divided into time periods called {\em re-configuration intervals}. 

\subsection{Constraints} \label{constraints}

\subsubsection{Service Delay} \label{service-delay-equation}
Service delay is defined as the time interval between the moment when an IoT node sends a service request and when it receives the response for that request. The average service delay of IoT nodes served by fog node $j$ for service $a$ is
\begin{IEEEeqnarray}{rCl}  \label{delay-app}
    &d_{aj} = [2d_{(I_a,j)}+w_{aj}+\frac{l^{rq}_a+l^{rp}_a}{r_{(I_a,j)}}]x_{aj}+\\
     &[2(d_{(I_a,j)}+d_{(j,k)})+w'_{ak}+(\frac{l^{rq}_a+l^{rp}_a}{r_{(I_a,j)}}+\frac{l^{rq}_a+l^{rp}_a}{r_{(j,k)}})](1-x_{aj}),\IEEEnonumber
\end{IEEEeqnarray}
where $k=h_a(j)$ is the index of a hosting cloud server for service $a$ (function $h_a(j)$ returns index of the cloud server to which the traffic for service $a$ is routed from fog node $j$). $I_a$ is the set of IoT nodes implementing (i.e. running) service $a$.

As can be inferred from the equation, when service $a$ is implemented on fog node $j$, ($x_{aj}=1$) service delay will be lower than when service $a$ is not implemented on fog node $j$ ($x_{aj}=0$). The equation for $d_{aj}$ is a conditional expression based on $x_{aj}$. The delay terms of each condition are propagation delay, waiting time (processing delay plus queueing delay), and transmission delay, in the order from left to right. $d_{(I_a,j)}$ and $r_{(I_a,j)}$ are the average propagation delay and average transmission rate between IoT nodes in $I_a$ and fog node $j$, respectively, and they are given as inputs to the \schemeName. The term $\frac{l^{rq}_a+l^{rp}_a}{r_{(I_a,j)}}$ captures the round-trip transmission delay, as it is the sum of transmission delays of the request ($\frac{l^{rq}_a}{r_{(I_a,j)}}$) and the response ($\frac{l^{rp}_a}{r_{(I_a,j)}}$) for the service $a$. Other variables are explained in Table \ref{parameters}. Multiple instances of a given service $a$ can be hosted in the cloud (on multiple cloud servers), as $h_a(.)$ can return the index of different cloud servers for different inputs $j$ and $j'$.

We use average/approximate values for the propagation delay and transmission rate between IoT nodes and fog nodes so that the Eq. (\ref{delay-app}) does not include indices of IoT nodes. We claim that using average/approximate values is reasonable, because, firstly, fog nodes are usually placed near IoT nodes, which means IoT nodes served by the same fog node have similar values of propagation delay and transmission rate. Secondly, if exact values were to be calculated, the FSC would need to have information about all the IoT nodes communicating with the fog nodes, which is not practical. 

As discussed above, $d_{(I_a,j)}$ and $r_{(I_a,j)}$ are given as inputs to the \schemeName. These values are either known by the IoT application owner and can be released to the FSP, or approximated by round-trip delay measurement techniques. If obtaining these values are not feasible in a scenario, we can change the scope of the definition of Eq. (\ref{delay-app}) to consider the service delay only within fog to cloud (see next section).
 
\subsubsection{Delay Budget} \label{delay-budget}
Note that the definition scope of Eq. (\ref{delay-app}) can be changed if obtaining average/approximate values for the propagation delay and transmission rate between IoT nodes and fog nodes is not feasible. This equation currently captures the IoT service delay, from the moment an IoT node sends a request and when it receives the response for that request. The scope of  Eq. (\ref{delay-app}) is currently IoT-fog-cloud (from IoT to cloud). We can change this scope to only fog-cloud (within fog and cloud). In other words, we can just capture the delay from the moment the request reaches a fog node. In this case $d_{aj}$ can be realized as the average {\em delay budget} for service $a$ at fog node $j$ within fog-cloud, and is equal to
\begin{equation} \label{service-delay-alt}
    d_{aj} = [w_{aj}]x_{aj}+
     [2d_{(j,k)}+w'_{ak}+\frac{l^{rq}_a+l^{rp}_a}{r_{(j,k)}}](1-x_{aj})
\end{equation}

Notice that the propagation and transmission delay components from IoT to fog are omitted from this equation. If this version of the equation is used, the delay threshold $th_a$ should also be reduced (and renamed to {\em threshold budget}) to consider only the threshold of the portion of delay within fog-cloud.

In the rest of this paper, we use the original Eq. (\ref{delay-app}) for the service delay. 
\subsubsection{Delay Violation}
To measure the quality of a given service $a$, we need to see what percentage of IoT requests do not meet the delay threshold $th_a$ (delay violation).  
We first need to check if average service delay of IoT nodes served by fog node $j$ for service $a$ (labeled as $d_{aj}$) is greater than the threshold $th_a$ defined in QoS requiremetns for service $a$. Let us define a binary variable $v_{aj}$ to indicate this:

\begin{equation} \label{qosv-binary}
v_{aj}=
    \begin{cases}
        1, & $if $d_{aj}>th_a\\
        0, & $otherwise$
    \end{cases},~~\forall j\in F, \forall a\in A.
\end{equation}

We define another variable that measures the delay violation of a given service according to the defined QoS requirements. Recall that the QoS requirement is that the percentage of delay samples from IoT nodes that exceed the delay threshold should be no more than $1-q_a$. We define $V^\%_a$ as the percentage of IoT service delay samples of service $a$ that do not meet the delay requirement. $V^\%_a$ can be calculated as follows 

\begin{equation} \label{qosv-percentage}
    V^\%_a = \frac{\sum_j{\lambda^{\text{in}}_{aj}v_{aj}}}{\sum_j{\lambda^{\text{in}}_{aj}}},~~~~~~\forall a\in A.
\end{equation}
Note that $V^\%_a$ is measured by the FSC, as a weighted average of $v_{aj}$, with $\lambda^{\text{in}}_{aj}$ as the weight. In practice, the FSC obtains $\lambda^{\text{in}}_{aj}$, the rate of incoming traffic to fog nodes, from the Traffic Monitor module.

We can now add the cost of delay violations to \textbf{P1}. Let $p_a$ denote the penalty that the FSP must pay per request of service $a$ if the delay requirements are violated by 1\%. The total cost of delay violation for the FSP will be
\begin{equation}
C^{\text{viol}}=\sum_{j\in F}{\sum_{a\in A}{\max(0,V^\%_a-(1-q_a))\times \lambda^{\text{in}}_{aj}p_a\tau_a}}.
\end{equation}

For instance if $q_a=97\%$, any violation percentage $V^\%_a$ greater than $3\%$ must be paid for by the FSP. Consider for a given service $a$, $q_a=97\%$, $p_a=4$, and $\lambda^{\text{in}}_{aj}=7$ for fog node $j$. If the violation percentage is $V^\%_a=5\%$, the FSP is charged the penalty of $(5-3)\times 7\times 4\times 6=336$ for one configuration interval of $\tau_a=6$ seconds for fog node $j$. The total penalty would be the sum of penalties for all services over all fog nodes.

Once we have discussed all the constraints of \textbf{P1}, we will add $C^{\text{viol}}$ to \textbf{P1} and rewrite the problem. Note that all of the unit cost parameters (e.g. $C^P_j$) are given as inputs to the \qdfsp~problem, hence are known to the FSP.

{\bf Remark}: It is worth noting that the above optimization problem aims to minimize the cost of the FSP, which may result in a solution that places some fog services far from their corresponding IoT devices. This can happen, for instance, if the number of IoT nodes that send requests to a particular fog node is small, or when the FSP's penalty for violating delay constraints is not high. In the former example, the few IoT nodes that send their requests to the less congested fog node may experience larger service delay if the service is not deployed on that fog node due to cost savings of not deploying the service. In the latter example, when the penalty for violating delay constraints is not high, \schemeName~may find a solution in which most services are deployed in the cloud. Nevertheless, if the fog service is delay-sensitive or has tight delay constraints, the penalty for violating delay constraints ($p_a$) will be set to a high value in the QoS agreement. 

\subsubsection{Resource Capacity}
Resource utilization of fog nodes and cloud servers shall not exceed their capacity, as it is formulated by 
\begin{IEEEeqnarray}{rCl}  
    &\sum_{a\in A}{x_{aj} L^S_a} < K^S_j, \text{and}~ \sum_{a\in A}{x_{aj} L^M_a} < K^M_j, ~~~~\forall j\in F,~~~~ \label{fog-storage-constraint}\\
    &\sum_{a\in A}{x'_{ak} L^S_a} < K'^S_k, \text{and}~ \sum_{a\in A}{x'_{ak} L^M_a} < K'^M_k, ~~~\forall k\in C.~~~~ \label{cloud-capacity-constraint}
\end{IEEEeqnarray}
If the ample capacity of the cloud is assumed to be unlimited, Eq. (\ref{cloud-capacity-constraint}) can be dropped. 
\subsubsection{Traffic Rates}
Different IoT requests have different processing times. In order to accurately evaluate the waiting times of fog nodes and cloud servers in our delay model, we need to account for the different processing times of the IoT requests. To this end, we express the incoming traffic rates to the fog nodes in units of instructions per unit time. 

Let $\Lambda_{aj}$ denote the arrival rate of instructions (in MIPS) to fog node $j$ for service $a$. This is the arrival rate of instructions of the incoming requests that are accepted for processing by fog node $j$ that is given by
\begin{equation} \label{fog-normal-traffic}
    \Lambda_{aj}={L^P_a \lambda^{\text{in}}_{aj} x_{aj}},~~~~~~~\forall j\in F, \forall a\in A.
 \end{equation}
 
The requests for different services have different processing times. Thus, in the definition of $\Lambda_{aj}$ we account for the different processing times by the inclusion of $L^P_a$. Note that if service $\tilde{a}$ is not deployed on fog node $j$, $\Lambda_{\tilde{a}j}=0$ since $x_{aj}=0$. In this case, the incoming traffic denoted by the rate $\lambda^{\text{in}}_{\tilde{a}j}$ will not be accepted by fog node $j$, and hence will be sent to the cloud. The rate of this ``rejected'' traffic is denoted by  
\begin{equation}
    \lambda^{\text{out}}_{aj} = \lambda^{\text{in}}_{aj}(1-x_{aj}),~~~~~~~~\forall j\in F, \forall a\in A.
\end{equation}
 Let $\lambda'^{\text{in}}_{ak}$ denote the incoming traffic rate to cloud server $k$ for service $a$. 
 Then we will have $\lambda'^{\text{in}}_{ak}=\sum_{j\in H_a^{-1}(k)}{\lambda^{\text{out}}_{aj}}$, where $H_a^{-1}(k)$ is a set of indices of all fog nodes that route the traffic for service $a$ to cloud server $k$ ($H_a^{-1}(k)=\{j|h_a(j)=k\}$). Similar to Eq. (\ref{fog-normal-traffic}), the arrival rate of instructions (in MIPS) to the cloud server $k$ for service $a$, $\Lambda'_{ak}$, can be obtained as
 \begin{equation}
    \Lambda'_{ak}=L^P_a\lambda'^{\text{in}}_{ak} x'_{ak},~~~~~~~\forall k\in C, \forall a\in A.
 \end{equation}

\subsubsection{Service Deployment in the Cloud}
If the incoming traffic rate to a cloud server for a particular service ($\lambda'^{\text{in}}_{ak}$) is $0$, the service could be safely released from the cloud server. This happens when the service is deployed on all the fog nodes that would otherwise route the traffic for that service to a cloud server. On the other hand, if at least one fog node routes traffic for that particular service to a cloud server ($\lambda^{\text{out}}_{aj}>0$), the service must not be released from the cloud server. This is because part of the traffic is being sent to and served by the cloud.
The following equation guarantees the above statement
 \begin{equation} \label{deployCloudServiceIfNeeded}
x'_{ak}=
    \begin{cases}
        0, & $if $\lambda'^{\text{in}}_{ak}=0\\
        1, & $if $\lambda'^{\text{in}}_{ak}>0
    \end{cases},~~~ \{k=h(j) | j\in F\}, \forall a\in A,
\end{equation}
which can be written as 
 \begin{equation} 
     \frac{\lambda'^{\text{in}}_{ak}}{W} \leq x'_{ak} \leq W\lambda'^{\text{in}}_{ak},~~~~~~~~ \{k=h(j) | j\in F\}, \forall a\in A,
 \end{equation}
 where $W$ is an arbitrary large number ($W>\max\limits_{a,k}(\lambda'^{\text{in}}_{ak})$).


\subsubsection{Waiting Times}
We have all the components of Eq. (\ref{delay-app}), except for the waiting times of fog nodes and cloud servers ($w_{aj}$ and $w'_{ak}$). We adopt a commonly used  M/M/$c$ queueing system \cite{m-m-c-1,m-m-c-2,m-m-c-3} model for a fog node with $n_j$ processing units, each with service rate $\mu_j$ and total arrival rate of $\sum_{a\in A}{\Lambda_{aj}}$ (total processing capacity of fog node $j$ will be $K^P_j=n_j\mu_j$). 

To model what fraction of processing units each service can obtain, we assume that the processing units of a fog node are allocated to the deployed services proportional to their processing needs ($L_a^P$). For instance, if the requests for service $a_1$ need twice the amount of processing than that of the requests for service $a_2$ ($L_{a_1}^P$ = $2\times L_{a_2}^P$), service $a_1$ should receive twice the service rate compared to service $a_2$. Correspondingly, we define $f_{aj}$, the fraction of service rate that service $a$ obtains at fog node $j$ as:
\begin{equation}
f_{aj}=
    \begin{cases}
        0, & $if $\sum_{a\in A}{x_{aj}}=0\\
        \frac{x_{aj}L_a^P}{\sum_{a\in A}{x_{aj}L_a^P}}, & $otherwise$
    \end{cases}.
\end{equation}
The first condition in the above equation is when no service is deployed on fog node $j$. Each deployed service can be seen as an M/M/$c$ queueing system with  service rate of $f_{aj}\times K^P_j=f_{aj}n_j\mu_j$, and arrival rate of $\Lambda_{aj}$ (in MIPS). Thus, the waiting time for requests of service $a$ at fog node $j$ will be \cite{bolch2006queueing}
\begin{equation} \label{proc-time-fog}
w_{aj} = \frac{1}{f_{aj}\mu_j} + \frac{\mathcal{P}^{Q}_{aj}}{f_{aj}K^P_j-\Lambda_{aj}},
\end{equation} 
where $\mathcal{P}^{Q}_{aj}$ is the probability that an arriving request to fog node $j$ for service $a$ has to wait in the queue. $\mathcal{P}^Q_{..}$ is also referred to as {\em Erlang's C formula} and is equal to 
\begin{equation} \label{erlangC}
\mathcal{P}^{Q}_{aj}=\frac{(n_j\rho_{aj})^{n_j}}{{n_j}!}\frac{\mathcal{P}^{0}_{aj}}{1-\rho_{aj}},
\end{equation} such that $\rho_{aj}=\frac{\Lambda_{aj}}{f_{aj}K^P_j}$ and
\begin{equation} \label{proc-time-fog-components}
\mathcal{P}^{0}_{aj}=\Big[ \sum_{c=0}^{n_j-1}{\frac{(n_j\rho_{aj})^c}{c!}}+\frac{(n_j\rho_{aj})^{n_j}}{{n_j}!}\frac{1}{1-\rho_{aj}} \Big]^{-1}.        
\end{equation}


 Note that in Eq. (\ref{proc-time-fog}) when a give service $a$ is not implemented on fog node $j$ (i.e. $f_{aj}=0$), waiting time $w_{aj}$ and $\rho_{aj}$ for that service are not defined.
  
Similarly, cloud server $k$ with $n'_k$ processing units (i.e. servers), each with service rate $\mu'_k$ and total arrival rate of $\sum_{a\in A}{\Lambda'_{ak}}$ can be seen as an M/M/$c$ queueing system (total processing capacity of cloud server $k$ will be $K'^P_k=n'_k\times \mu'_k$). Therefore, similar to Eq. (\ref{proc-time-fog}), $w'_{ak}$, the waiting time for requests of service $a$ at cloud server $k$, could be derived as 
 \begin{equation} \label{proc-time-cloud}
    w'_{ak} = \frac{1}{f'_{ak}\mu'_k} + \frac{\mathcal{P}'^{Q}_{ak}}{f'_{ak}K'^P_k-\Lambda'_{ak}}, ~~~~~~~~\forall a\in A, \forall k\in C.
 \end{equation}
 An equation similar to Eq. (\ref{erlangC}) is defined for $\mathcal{P}'^{Q}_{ak}$, the probability of queueing at cloud server $k$. Note that for simplicity, instead of modeling each cloud server a M/M/$c$ queue, one may also model the whole cloud as an M/M/$\infty$ queueing system. Finally, stability constraints of the M/M/$c$ queues for the services on fog nodes and cloud servers imply
\begin{IEEEeqnarray}{rCl} 
     \Lambda_{aj} < f_{aj}K^P_j,~~~~\forall a\in A, \forall j\in F.\\
     \Lambda'_{ak} < f'_{ak}K'^P_k,~~\forall a\in A, \forall k\in C.
\end{IEEEeqnarray}
\subsection{Final Optimization Formulation}
\textbf{P1} can be rewritten as \textbf{P2} with the same constraints:
\begin{IEEEeqnarray}{rCl} 
     \textbf{P2}: &\text{Minimize}& (C^{\text{proc}}_{C}+C^{\text{proc}}_{F})+(C^{\text{stor}}_{C}+C^{\text{stor}}_{F})+\IEEEnonumber\\ 
     &~~&(C^{\text{comm}}_{FC}+C^{\text{comm}}_{FF}+C^{\text{depl}}_{\Phi F})+(C^{\text{viol}})\IEEEnonumber\\ 
    &\text{~Subject~to}&\text{~equations~} (1),(2),(15)-(28). \IEEEnonumber
\end{IEEEeqnarray}

The objective function of \textbf{P2} is the summation of eight cost functions. In some scenarios, though, it is possible that certain costs (e.g. cost of storage) are the dominant factors in this summation. In order to consider the general problem in this paper, we consider all of the eight cost functions; however, some of them can be omitted in certain scenarios if needed. Note that in this optimization problem, we only consider the costs of fog-to-fog and fog-to-cloud communication. In other words, the cost of communication between IoT and fog is not considered in the optimization problem, since this is usually outside the control of the FSP. 

Since the incoming traffic to fog nodes is changing over time, to address the \qdfsp, the FSC needs to dynamically adjust the provisioning of services over time to meet the QoS requirements. To do this, one approach is to solve the optimization problem periodically, which may not be feasible for large network due to non-scalability of INLP. Another alternative is using some incremental algorithms that are more efficient than solving the optimization problem. In the next section, we propose two such algorithms.

\section{\schemeName's Greedy Algorithms} \label{Heuristic}
In this section, we describe our proposed greedy algorithms that are called periodically to efficiently address the \qdfsp~problem. We propose two algorithms: Min-Viol, which aims to minimize the delay violations, and Min-Cost, whose goal is minimizing the total cost. The Min-Viol and Min-Cost are shown in Algorithm 1 and Algorithm 2 listings, respectively. Both algorithms try to address the \qdfsp~problem efficiently, and if implemented, will be periodically run by the FSC every $\tau_a$ seconds. The proposed algorithms are discussed in more detail in the following subsection. 
\begin{algorithm} 
\caption{\schemeName~Min-Viol}
\begin{algorithmic}[1]
\renewcommand{\algorithmicrequire}{\textbf{Input:}}
\renewcommand{\algorithmicensure}{\textbf{Output:}}
\Require Service $a$ size stats, $G$, $q_a$, $th_a$, $\lambda^{\text{in}}_{aj}$, cost parameters 
\Ensure Placement of service $a$ that minimizes delay violation
\State Deploy/release service $a$ in cloud according to Eq. (\ref{deployCloudServiceIfNeeded})
\State Read service $a$'s incoming traffic rate to fog nodes 
\State List $L\leftarrow$ sort fog nodes in descending order of traffic rate
\State \Call{calcViolPerc}{$x_{aj}$, $\lambda^{\text{in}}_{aj}$, $th_a$} \Comment{$V^\%_a$ updated}
\State \textit{canDeploy} = true
\While{\textit{canDeploy} and ($V^\%_a > 1-q_a$)}
\Comment{Deploy}
\State $j$ = remove from the list $L$ index of the next fog node
\If{L is empty}
\State \textit{canDeploy} = false
\EndIf
\If{(service $a$ is not deployed on fog node $j$) and (fog node $j$ has enough storage and memory)}
\State Deploy service $a$ on fog node $j$ \Comment{$x_{aj}=1$}
\State \Call{calcViolPerc}{$x_{aj}$, $\lambda^{\text{in}}_{aj}$, $th_a$} \Comment{update $V^\%_a$}
\EndIf
\EndWhile
\State \textit{canRelease} = true
\While{canRelease} \Comment{Release}
\State $j$ = get from the tail of $L$ a fog node $j$, on which service $a$ is deployed
\State Set $x_{aj}=0$ (and set $x'_{ah_a(j)}=1$, if $x'_{ah_a(j)}=0$)
\State \Call{calcViolPerc}{$x_{aj}$, $\lambda^{\text{in}}_{aj}$, $th_a$} \Comment{update $V^\%_a$}
\If{$V^\%_a \leq 1-q_a$}
\State Release service $a$ on fog node $j$ \Comment{$x_{aj}=0$}
\Else \Comment{if releasing would cause violation}
\State Set $x_{aj}=1$, and \textit{canRelease} = false \Comment{exit}
\EndIf
\EndWhile
\end{algorithmic}
\end{algorithm}
\subsection{Description}
\subsubsection{Min-Viol}
The high-level rationale behind the Min-Viol algorithm is to deploy on the fog nodes those services for which there is high demand, and to release from the fog nodes services for which there is low demand, while keeping the violation low, as per the QoS requirements.
Note that when a given service $a$ is not deployed on a given fog node $j$, the traffic routed to the cloud for service $a$ might pass through the fog node $j$ (or the monitoring-enabled device to which fog node $j$ is connected). This can happen, for instance, when the demand for a service $a$ in one location increases and many requests for (currently cloud-hosted) service $a$ pass through fog node $j$. Thus, in \schemeName, fog nodes monitor the traffic for different services and communicate with FSC's Traffic Monitor module to act on such demands. 

First, the incoming traffic rates to the fog nodes are read from the Traffic Monitor module, based on which the fog nodes are sorted in descending order (lines 2-3). Next, using the method $\Call{calcViolPerc}{.}$ (shown in Algorithm 3), the percentage of the violating IoT requests for service $a$ ($V^\%_a$) is calculated. As long as this percentage is more than $1-q_a$ (lines 6-15), the algorithm keeps deploying services on the fog nodes, and once $V^\%_a \leq 1-q_a$, it exits the loop. The boolean variable $canDeploy$ is for eliminating blocking of the while loop ($canDeploy$ becomes false when $V^\%_a$ does not drop below $1-q_a$ after going through the list). Note that the services are deployed first on the fog nodes with higher incoming traffic rate, so that $V^\%_a$ decreases faster.

When $V^\%_a \leq 1-q_a$, we might still be able to release services on the fog nodes with small incoming traffic rate, without violating QoS requirements. The second loop (lines 17-26) releases services on the fog nodes with smaller incoming traffic rate. First, we try to find a fog node with small incoming traffic rate (from the end of sorted list $L$) and check if releasing the already-deployed service would cause any delay violations (lines 19-20). If releasing does not cause violation (line 21), we go ahead and release the service (line 22); otherwise, we do not release the service and exit (line 24), because releasing service on the next fog nodes with higher incoming traffic would cause even more delay violations. Min-Viol requires service $a$'s average size statistics (i.e., service storage size, size of request and reply, amount of processing), graph $G$, $q_a$, and $th_a$ as input.

In both Algorithm 1 and Algorithm 2, step 1 is useful for deployment settings where the FSC has access to the cloud resources as well and is able to deploy services on the cloud servers. If the FSC has access to the cloud resources, an instance of service $a$ is deployed or released in the cloud according to Eq. (\ref{deployCloudServiceIfNeeded}). If the FSC does not have access to the cloud resources, an instance of service $a$ should be always running in the cloud, as the cloud will be the last resort for the requests not processed by the fog nodes.
\begin{algorithm} 
\caption{\schemeName~Min-Cost}
\begin{algorithmic}[1]
\renewcommand{\algorithmicrequire}{\textbf{Input:}}
\renewcommand{\algorithmicensure}{\textbf{Output:}}
\Require Service $a$ size stats, $G$, $q_a$, $th_a$, $\lambda^{\text{in}}_{aj}$, cost parameters 
\Ensure Placement of service $a$ that minimizes cost
\State Deploy/release service $a$ in cloud according to Eq. (\ref{deployCloudServiceIfNeeded})
\State Read service $a$'s incoming traffic rate to fog nodes 
\State List $L\leftarrow$ sort fog nodes in descending order of traffic rate

\ForAll{fog node $j$ in $L$} \Comment{Deploy}

\If{fog node $j$ has enough storage and memory}
\If{cost savings of deploying $a$ on $j$ $>$ expenses of deploying service $a$ on $j$}
\State Deploy service $a$ on fog node $j$ \Comment{$x_{aj}=1$}
\State \Call{calcViolPerc}{$x_{aj}$, $\lambda^{\text{in}}_{aj}$, $th_a$} \Comment{update $V^\%_a$}
\EndIf
\EndIf
\EndFor
\ForAll{fog node $j$ in $L$(reverse)} \Comment{Release}
\If{cost savings of releasing $a$ on $j$ $>$ expenses of releasing service $a$ on $j$}
\State Release service $a$ on fog node $j$ \Comment{$x_{aj}=0$}
\State \Call{calcViolPerc}{$x_{aj}$, $\lambda^{\text{in}}_{aj}$, $th_a$} \Comment{update $V^\%_a$}
\EndIf
\EndFor
\end{algorithmic}
\end{algorithm}
\subsubsection{Min-Cost}
The main idea of the Min-Cost algorithm is similar to that of Min-Viol; however, the major concern in Min-Cost is minimizing the cost. Min-Cost tries to minimize the cost by checking if deploying or releasing services will increase the revenue (or equally will decrease cost). Similar to the Min-Viol algorithm, the incoming traffic rates to fog nodes are read from the Traffic Monitor module, based on which the fog nodes are sorted in descending order (lines 2-3). Next, we iterate through fog nodes and check if deploying a service make sense, in terms of minimizing the cost (lines 4-11). Line 6 checks if the cost savings of deploying service $a$ on fog node $j$, is larger than the expenses (or losses) when the service $a$ is not deployed on fog node $j$. The cost savings of deploying a service are due to the reduced cost of communication between fog and cloud, the reduced costs of storage and processing in the cloud, and the (possible) reduced cost of delay violations when the service $a$ is implemented on fog node $j$. Conversely, the expenses are the cost of service deployment and the increased costs of storage and processing in the fog. All the mentioned costs are calculated and compared for the duration of one interval ($\tau_a$).

\begin{algorithm}
\caption{Calculate Delay Violation Percentage}
\begin{algorithmic}[1]
\renewcommand{\algorithmicrequire}{\textbf{Input:}}
\renewcommand{\algorithmicensure}{\textbf{Output:}}
\Require $x_{aj}$, $\lambda^{\text{in}}_{aj}$, $th_a$
\Ensure Percentage of IoT requests that do not meet the delay requirement for service $a$ ($V^\%_a$)
\Procedure{calcViolPerc}{$x_{aj}$, $\lambda^{\text{in}}_{aj}$, $th_a$}
\State Calculate $d_{aj}$ for $a$ for all fog nodes (Eq. (\ref{delay-app}) or (\ref{service-delay-alt}))
\State Calculate $v_{aj}$ for $a$ for all fog nodes (Eq. (\ref{qosv-binary}))
\State Calculate $V^\%_a$ using Eq. (\ref{qosv-percentage})
\State \textbf{return} $V^\%_a$
\EndProcedure
\end{algorithmic}
\end{algorithm}

Similar to deploying (lines 4-11), for releasing, we iterate through fog nodes, in increasing order of their incoming traffic rates and check if releasing a service make sense, in terms of minimizing the cost (lines 12-17). Likewise, line 13 checks if the cost savings of releasing service $a$ on fog node $j$ is greater than the expenses when the service $a$ is deployed on fog node $j$. The cost savings of releasing a service are due to the reduced costs of storage and processing in the fog when the service is released, while the expenses are due to the increased cost of communication between fog and cloud, the increased costs of storage and processing in the cloud, and the (possible) increased cost of delay violations. Min-Cost's inputs are similar to those of Min-Viol; additionally, Min-Cost requires the unit cost parameters (processing, storage, and networking) to calculate the cost using equations (3)-(9).
\subsection{Complexity} \label{complexity}
The time complexity of the proposed greedy algorithms shown in Algorithm 1 and Algorithm 2 listings are discussed below. 
\subsubsection{Min-Viol}
The asymptotic complexity of lines 1-3 is $O(|F|\log |F|)$ due to the sorting of fog nodes. The complexity of lines 4-5 is $O(|A||F|)$, since function $\Call{calcViolPerc}{.}$ calculates $d_{aj}$ for all fog nodes. The steps in lines 6-15 run at most $|F|$ times, hence their complexity is $O(|A||F|^2)$. Similarly, the complexity of lines 17-26 is $O(|A||F|^2)$, therefore, the overall complexity of Min-Viol for one service is $O(|A||F|^2)$.\subsubsection{Min-Cost}
Similar to the above analysis, the asymptotic complexity of lines 1-3, lines 4-11, and lines 12-17 is $O(|F|\log |F|)$, $O(|A||F|^2)$, and $O(|A||F|^2)$, respectively. The overall complexity of Min-Cost for one service is $O(|A||F|^2)$. We can see both algorithms have the same asymptotic complexity. In the next section, we will compare the algorithms numerically using our extensive simulations. 
\section{Numerical Evaluation} \label{results}
In this section, we discuss the settings and results of the numerical evaluation of the proposed framework. Our numerical evaluations are performed using a simulation environment written in Java\footnote{{\scriptsize Available at https://github.com/ashkan-software/\schemNameUnformatted-simulator}} and are based on real-world traffic traces. 

\subsection{Simulation Settings}
\subsubsection{Simulation Environment}
The evaluation is performed using a Java program that simulates a network of fog nodes, IoT nodes, and cloud servers. The simulator can either read the traffic trace files or can generate arbitrary traffic patterns based on a Discrete Time Markov Chain (DTMC). In either case, the simulator solves the optimization problem and/or runs our proposed greedy algorithms. The parameters of the simulation are summarized in Table \ref{sim-parameters}, and are explained in what follows. 
\subsubsection{Experiments}
To fully understand the benefits of our proposed scheme, we have conducted five experiments to study the impact of the different parameters and factors of the framework that affect the results. The purpose of the experiments and their settings are summarized in Table \ref{sim-experiments} and are briefly explained here. Experiment-1 (results are shown in the left column of Fig. \ref{sim-results-1}) is to study the benefits of our proposed greedy algorithms under a real-world traffic trace. In experiment-2 (results are shown in the right column of Fig. \ref{sim-results-1}) we compare the performance of our proposed greedy algorithms with that of the optimal solution achieved using the optimization problem. Experiment-3 (results are shown in the left column of Fig. \ref{sim-results-2}) is for investigating the impact of the delay threshold, $th_a$, whereas experiment-4 (results are shown in the right column of Fig. \ref{sim-results-2}) is for investigating the impact of the configuration interval, $\tau_a$. Finally, in experiment-5 (results are shown in Table \ref{scalability}) we analyze the scalability of our proposed greedy algorithms using larger network topologies and more services. 

\subsubsection{Topology}
The logical topology of the network for the first four experiments are summarized in Table \ref{sim-experiments}. The topology of the fifth experiment is not fixed and will be discussed later. The fog nodes are paired with different cloud servers initially according to a random uniform distribution. The number of IoT nodes is not required for our experiments, as we use the aggregated incoming traffic rate from the IoT nodes to the fog nodes. 

The propagation delay can be estimated by halving the round-trip time of small packets; and as evaluated in our previous study \cite{ashkan-fog-delay}; it is assumed to be $U(1,2)$ ms between the IoT nodes and the fog nodes, and $U(15,35)$ ms between the fog nodes and the cloud servers. The transmission medium between the IoT nodes and the fog nodes is assumed to be either (50\% chance for each case) one hop WiFi, or WiFi and a 1 Gbps Ethernet link. The fog nodes and the cloud servers are assumed to be 6-10 hops apart, while their communication path consists of 10 Gbps (arbitrary) and 100 Gbps (up to 2) links. The transmission rate of the link between the FSC and the fog nodes is assumed to be 10 Gbps. 

\begin{table}[!t]
    \caption{Simulation Parameters. $U(a,b)$ indicates random uniform distribution between $a$ and $b$. ``MI'' is Million Instructions.}
    \centering
    {\footnotesize
        {\renewcommand\arraystretch{1.0} 
            \begin{tabular}{|lr|lr|}
                \hline
                $q_a$ & $U(90,99.999)\%$ & $u_e$ & 0.2 per Gb\\
                 $th_a$ & $10$ ms & $u_{(\Phi,j)}$ & 0.5 per Gb\\
                 $d_{(I_a,j)}$ & $U(1,2)$ ms & $l^{rq}_a$ & $U(10, 26)$ KB\\
                 $d_{(j,k)}$ & $U(15,35)$ ms & $l^{rp}_a$& $U(10, 20)$ B \\
             Core link & 10 Gbps, 100 Gbps & $L_a^P$&$U(50,200)$ MI per req\\
             Edge link & 54 Mbps, 1 Gbps & $L_a^S$ & $U(50,500)$ MB\\
                $K^P_j$ & $U(800,1300)$ MIPS & $L_a^M$ & $U(2,400)$ MB \\
                $K'^P_k$ & $U(16K,26K)$ MIPS & $p_a$ & $U(2,5)$ per req per \%\\
                $K^M_j$ & 8 GB & $C^P_j$ & 0.002 per MI\\
                $K'^M_k$ & 32 GB & $C'^P_k$ & 0.002 per MI\\
                $K^S_j$, $K'^S_k$  & $\geq$25, $\geq$250 GB & $C^S_j$ & 0.004 per Gb per sec\\
                $n_j$, $n'_k$ & 4, 8 Proc. Units & $C'^S_k$ & 0.004 per Gb per sec\\
                \hline
            \end{tabular}
        }}
        \label{sim-parameters}
    \end{table}

\subsubsection{QoS parameters}The level of quality of service of different services is assumed to be a uniform random number between {\em loose}$=90\%$ and {\em strict}$=99.999\%$. Since in this paper we focus on delay-sensitive fog services, we set the penalty of violating the delay threshold to a random number in $U(10,20)$ per \% per sec for experiment-1 and experiment-2, and in $U(100,200)$ per \% per sec for the other experiments.
\subsubsection{Real-world Traffic Traces}
In order to evaluate our framework and obtain realistic results, we have employed real-world traffic traces, taken from MAWI Working Group traffic archive \cite{wide}. MAWI's traffic archive is maintained by daily trace captures at the transit link of WIDE to their upstream ISP. We have used the traces of {\em 2017/04/12-13} in this paper for modeling the incoming traffic rates to fog nodes from IoT nodes. The traffic traces are provided by MAWI Working Group in chunks of 15-minute intervals. 

To model the cloud, we have chosen a particular class B subnet in the traffic traces to represent the IP addresses of the FSP's cloud servers. For modeling the fog nodes, we have selected 10 regions/cities (i.e. subnets) from the traffic trace file with a large number of packets and assumed that there is one fog node in each city. The fog node can later serve the IoT requests coming from the city where it resides. We assumed that the packets destined to or originated from a class C subnet belong to the same region/city. 

We consider the TCP and UDP packets in the traffic trace to account for the requests generated by the IoT nodes, because common IoT protocols, such as MQTT, COAP, mDNS, and AMQP, use TCP and UDP to carry their messages \cite{iot-survey}. The tuple $<$IP address, port number$>$ (destination) represents a particular fog/cloud service, to which IoT requests are sent. 

\subsubsection{DTMC-based Traffic Traces}
As mentioned before, we have also developed a traffic trace generator based on a Discrete Time Markov Chain (DTMC), as the number of regions/cities representing fog nodes is limited in the real-world MAWI traffic traces. DTMCs can be used to simulate various traffic traces by considering a Markov process for the traffic rate. Essentially, the states in the Markov process represent different traffic rates (quantized), and the transition probabilities are the fraction of time that traffic rate changes from a particular rate to a new rate. 

In our simulation, the DTMC-based traffic trace generator has 30 states and is constructed based on the 48-hour trace of {\em 2017/04/12-13}. The generated traffic trace is random and each run results in a new traffic trace. Our obtained results based on such trace produced by DTMC-based traffic trace generator are discussed in experiment-4 and experiment-5. 

\begin{table}[!t]
\caption{Simulation Experiments Settings}
\newcolumntype{L}[1]{>{\raggedright\let\newline\\\arraybackslash\hspace{0pt}}m{#1}}
    \centering
    {\footnotesize
        {\renewcommand\arraystretch{1.0} 
            \begin{tabular}{|L{0.03\columnwidth}|L{0.35\columnwidth}|L{0.21\columnwidth}|L{0.22\columnwidth}|}
                \hline
                \rotatebox{90}{Experiment~} & Purpose of Experiment & Traffic Trace & Topology Settings\\
                \hline
                1 & Study the benefits of Min-Cost and Min-Viol under real-world traffic trace & 48-hour trace (2017/04/12-13) & 3 cloud servers, 10 fog nodes, ~~40 services\\
                2 & Compare Min-Cost and Min-Viol greedy algorithms with the optimal & 2-hour trace ((12:00PM-2:00PM) of 2017/04/12) & 1 cloud server, 10 fog nodes, ~~2 services\\
                3 & Study the impact of delay threshold $th_a$ on the proposed algorithms & 4-hour trace ((12:00PM-4:00PM) of 2017/04/12) & 3 cloud servers, 10 fog nodes, ~~20 services\\
                4 & Study the impact of reconfiguration interval length $\tau_a$ on the proposed algorithms & DTMC generated traffic trace & 3 cloud servers, 15 fog nodes, ~~~50 services\\
                
                \hline
            \end{tabular}
        }}
        \label{sim-experiments}
    \end{table}
    
\subsubsection{Capacity}
In the simulation, the processing capacity of each fog node, $K^P_j$, is $U(800,1300)$ MIPS \cite{cooperative}, and the processing capacity of each cloud server, $K'^P_k$, is assumed to be 20 times that of the fog nodes. The storage capacity of the fog nodes, $K^S_j$, is assumed to be more than 25 GB, to host at most 50 services of the maximum size (size of services, $L_a^S$, is $U(50,500)$ MB for typical Linux containers). The storage capacity of the cloud servers is 10 times that of the fog nodes. The memory capacity of the fog nodes and the cloud servers, $K^M_j$ and $K'^M_k$, is 8 GB and 32 GB, respectively. Fog nodes assumed to have 4 processing units whereas cloud servers assumed to have 8. These capacity assumptions are reasonable with regards to the current cloud computing instance sizes (e.g., 32 GB memory is the default config for Amazon AWS t2.2xlarge, m5d.2xlarge, m4.2xlarge, and h1.2xlarge instance types\footnote{{\scriptsize See https://aws.amazon.com/ec2/instance-types/}}). 

\subsubsection{App Size}
Several applications could be used to show the applicability and benefits of the proposed framework. In order to obtain realistic values for different parameters of the framework and show the benefits of the scheme under practical situations, we consider mobile augmented reality (MAR) as the application that runs on IoT devices. In MAR applications, the requests have a delay threshold of 10 ms \cite{byers2017architectural}. The average size of request and response of the MAR applications are set as $U(10,26)$ KB and $U(10,20)$ byte, respectively \cite{impact}, and the required amount of processing for the services is $U(50,200)$ MI per request \cite{skarlat2017towards}.

\subsubsection{Deploy and Release Delay}
One question that we need to answer is: how much delay does deploying or releasing a service incur? Is this delay negligible in practical fog networks? If deploying or releasing services fog service causes extra delay, this could have a significant impact on QoS. Nonetheless, deploying and releasing containers takes less than 50 ms \cite{container-as-service}, wheres the interval of monitoring traffic and running the \schemeName~for deploying services in a real-world setting would be in the order of tens of seconds to minutes. In our simulations, the start-up delay of the service containers is set to 50 ms.  

\subsubsection{Cost}
The cost of communication between nodes, $u_e$, is $0.2$ per Gb, and between the FSC and fog nodes, $u_{(\Phi,j)}$, is $0.5$ per Gb. At the time of writeup of this paper, there was no standard fog pricing, by which we can define reference costs. The cost of processing in fog nodes and cloud servers is set to 0.002 per MI. The cost of storage in fog nodes and cloud servers is 0.004 per Gb per second. 

\subsection{Results}

The results of the simulation are shown in Fig. \ref{sim-results-1}, Fig. \ref{sim-results-2}, and Table \ref{scalability}. We will discuss the five experiments in more detail in this subsection. In all figures, the label ``All Cloud'' indicates a setting where the IoT requests are sent directly to the cloud (that is, fog nodes do not process the IoT requests). ``Min-Viol'' and ``Min-Cost'' represent the two proposed greedy algorithms, and ``Static Fog'' is a baseline technique where the services are deployed statically at the beginning, as opposed to the dynamic deployment. Static Fog uses the Min-Cost algorithm, and finds a one-time placement of the fog services at the beginning of the run, using the average traffic rates of the fog nodes as the input. The ``Optimal'' label indicates a setting where (optimal) results are obtained using the optimization problem introduced in Section \ref{QDFSP-INLP} (equations $(1)-(28)$). 
\subsubsection{Experiment-1}
\begin{figure}[!t]
    \begin{minipage}{.48\linewidth}
    \begin{tabular}{l}
        \subfloat[{\scriptsize Traffic Trace}]{\includegraphics[width=0.97\linewidth]{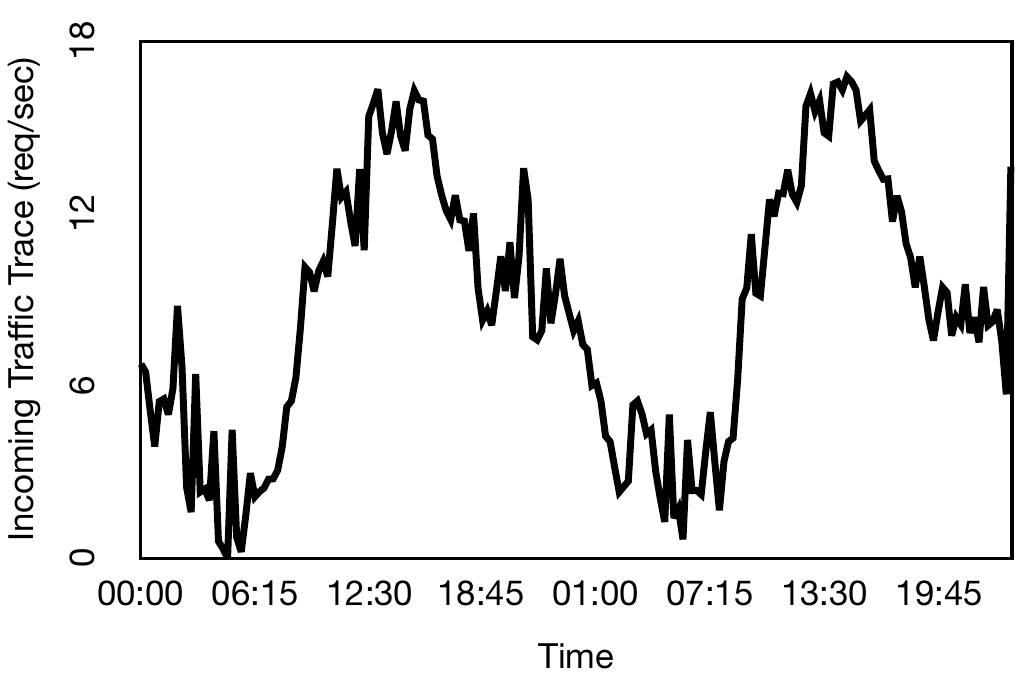}
        \label{trace1}}\\
        \subfloat[{\scriptsize Average Service Delay}]{\includegraphics[width=0.99\linewidth]{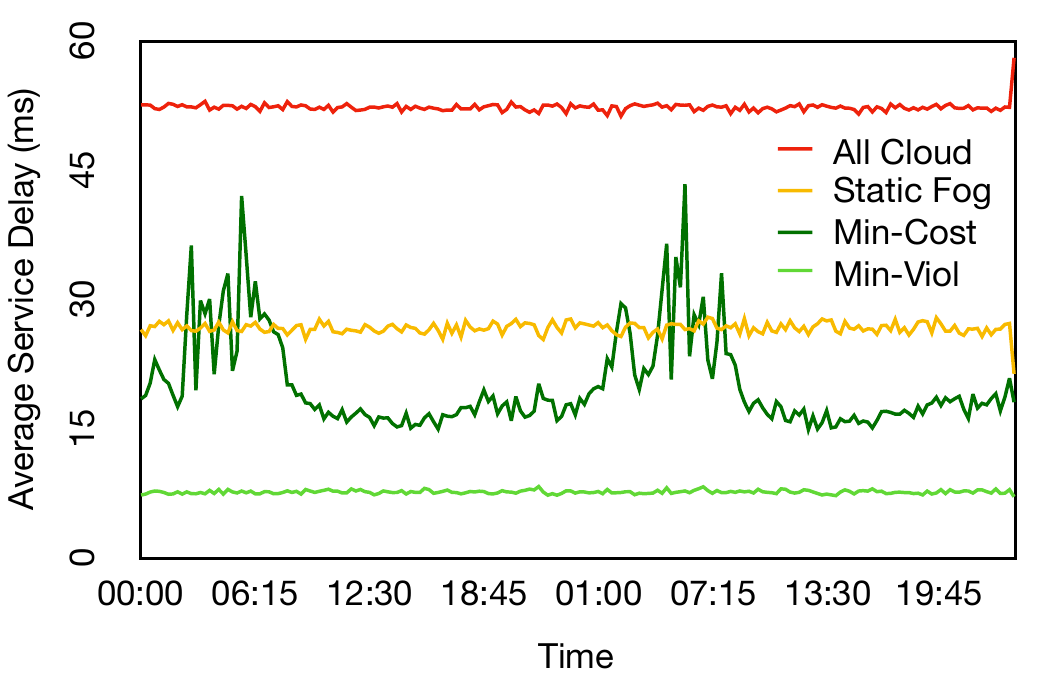}
        \label{trace1-delay}}\\
        \subfloat[{\scriptsize Average Cost}]{\includegraphics[width=0.97\linewidth]{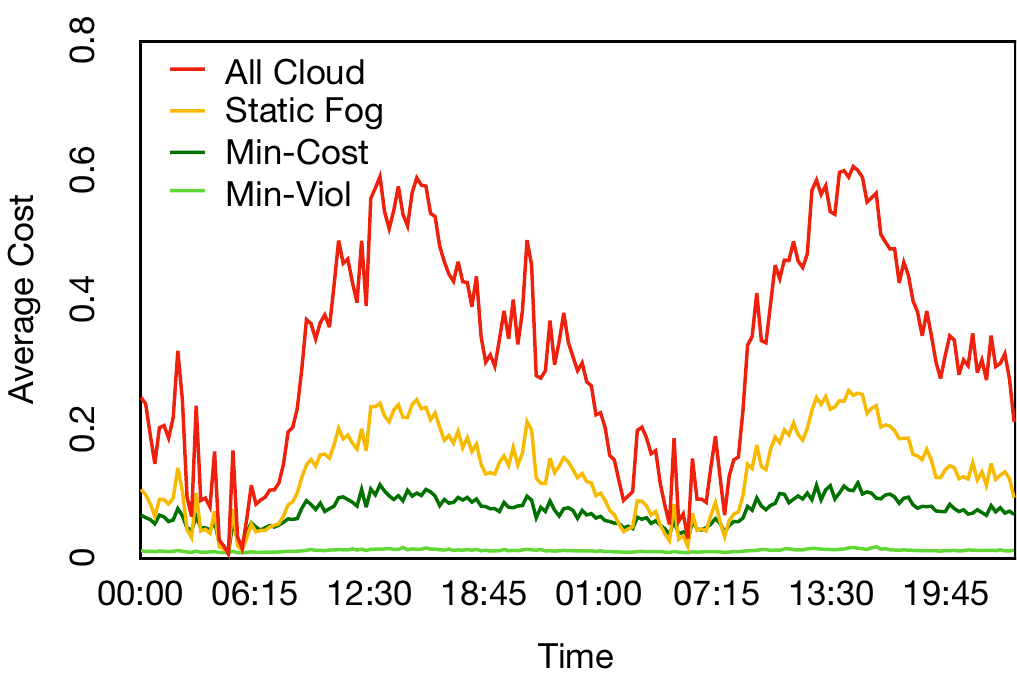}
        \label{trace1-cost}}\\
        \subfloat[{\scriptsize Average Delay Violation}]{\includegraphics[width=0.99\linewidth]{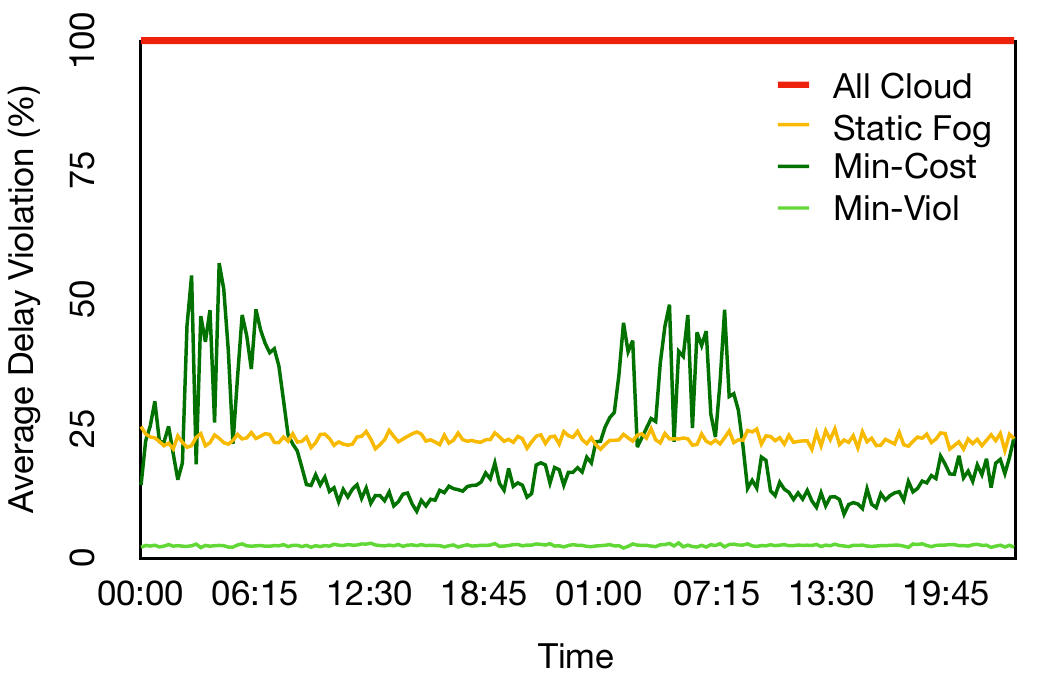}
        \label{trace1-viol}}\\
        \subfloat[{\scriptsize Average Number of Fog Services}]{\includegraphics[width=0.99\linewidth]{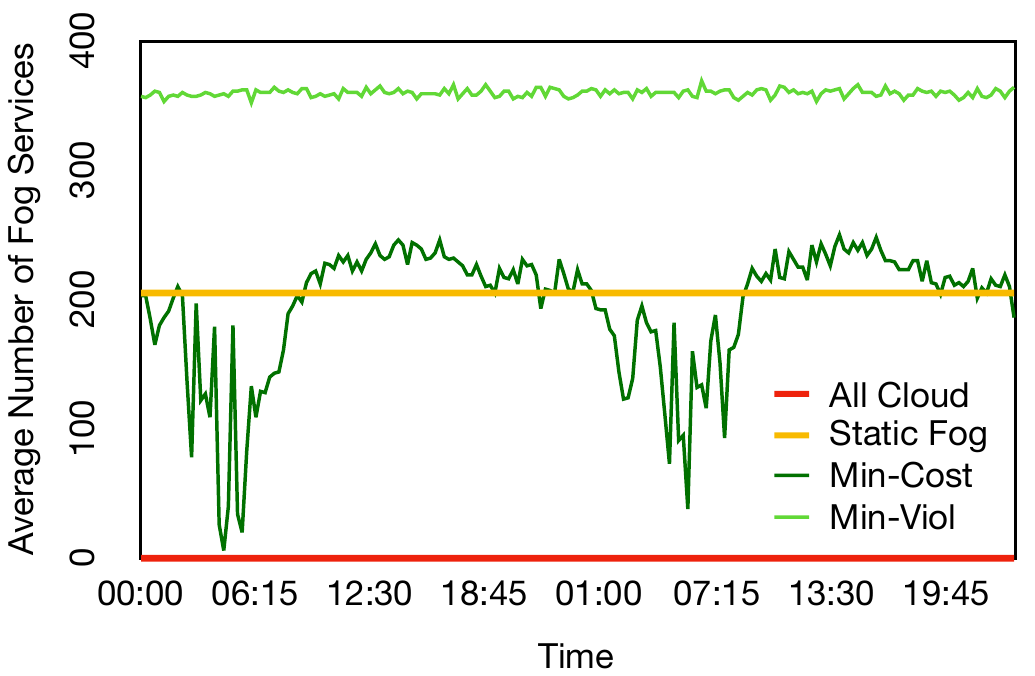}
        \label{trace1-services}}\\
        \subfloat[{\scriptsize Average Number of Cloud Services}]{\includegraphics[width=\linewidth]{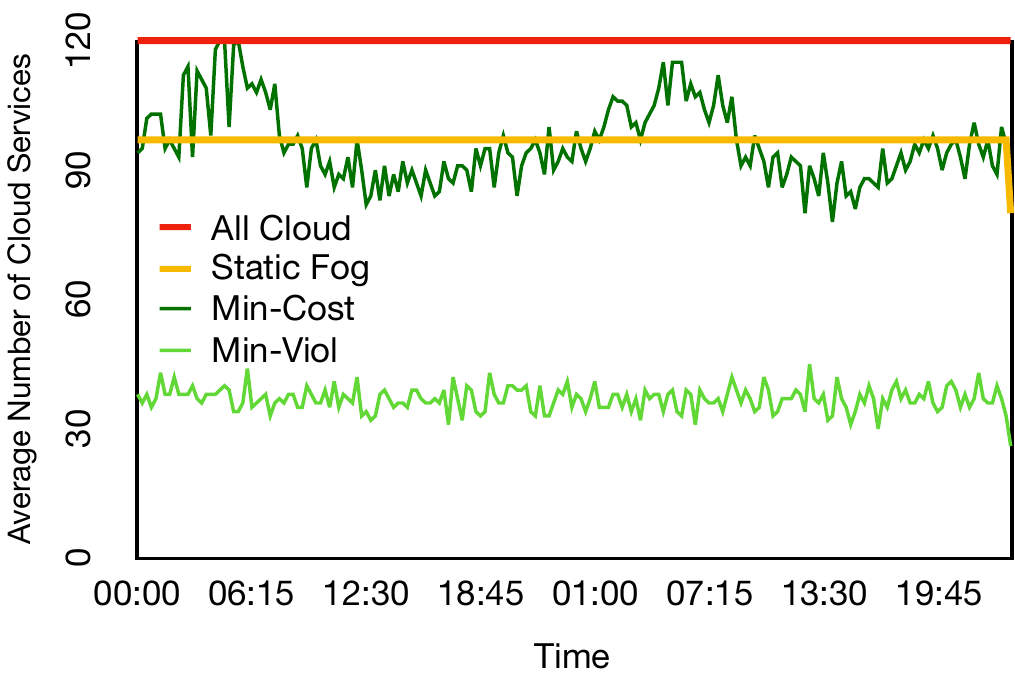}
        \label{trace1-services-cloud}}
        
    \end{tabular}
    \end{minipage}%
    ~
    \begin{minipage}{.48\linewidth}
    \begin{tabular}{l}
        \subfloat[{\scriptsize Traffic Trace}]{\includegraphics[width=0.98\linewidth]{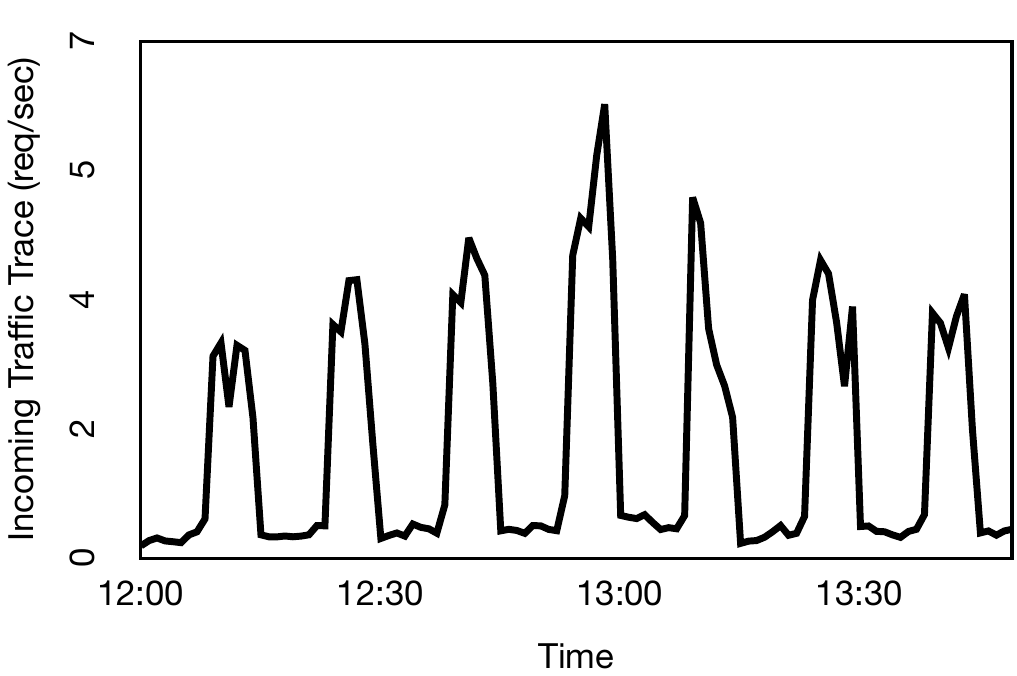}
        \label{trace2}}\\
        \subfloat[{\scriptsize Average Service Delay}]{\includegraphics[width=1.01\linewidth]{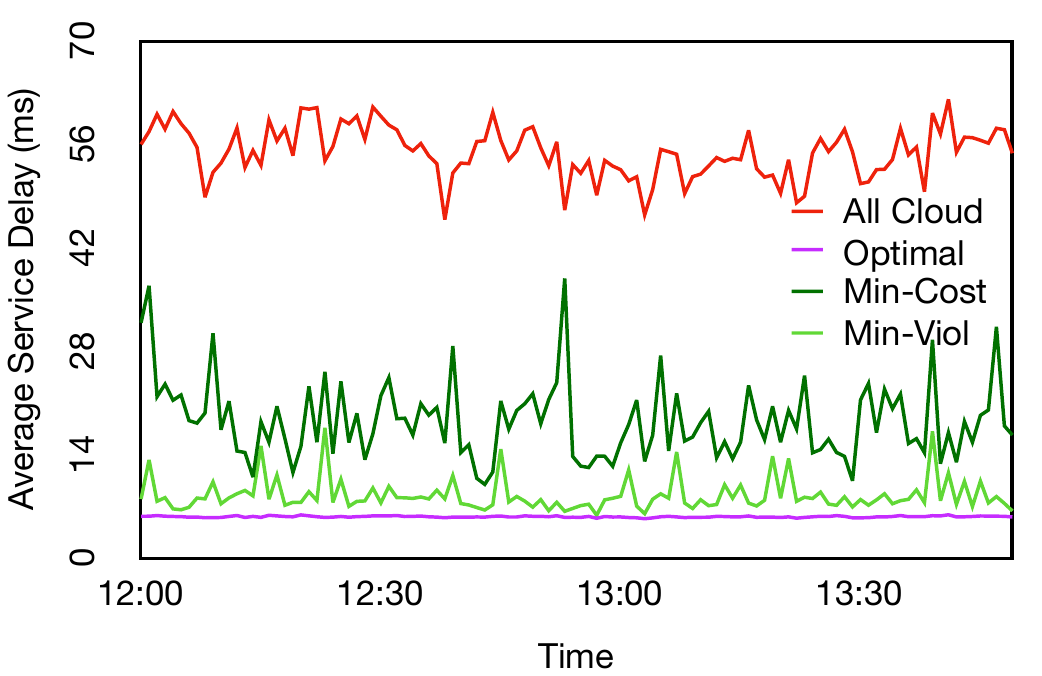}
        \label{trace2-delay}}\\
        \subfloat[{\scriptsize Average Cost}]{\includegraphics[width=0.98\linewidth]{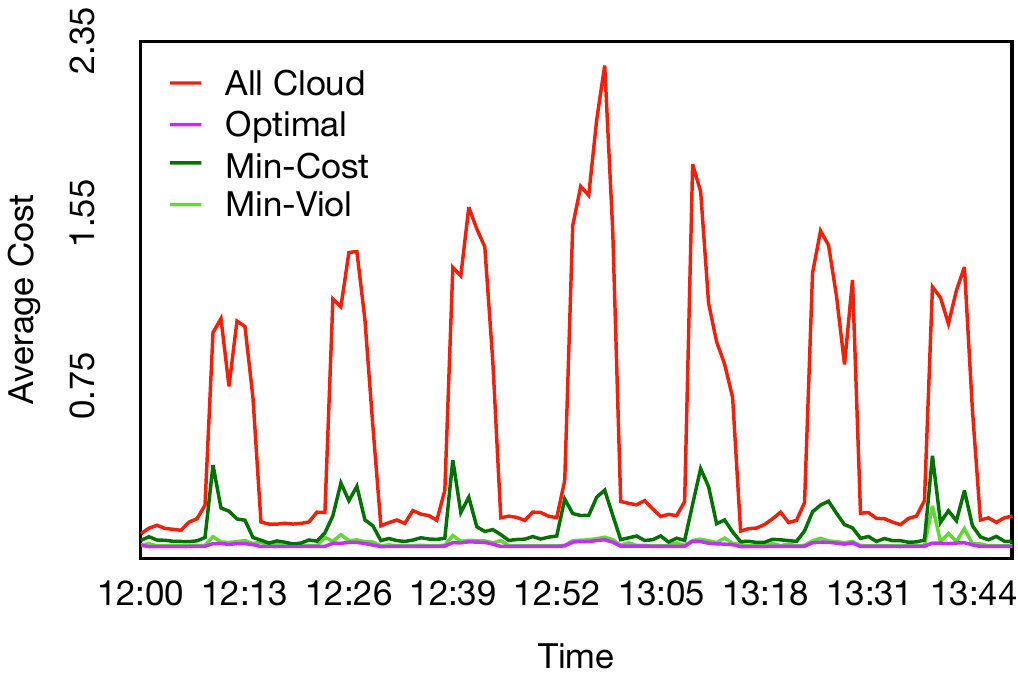}
        \label{trace2-cost}}\\
        \subfloat[{\scriptsize Average Delay Violation}]{\includegraphics[width=1.01\linewidth]{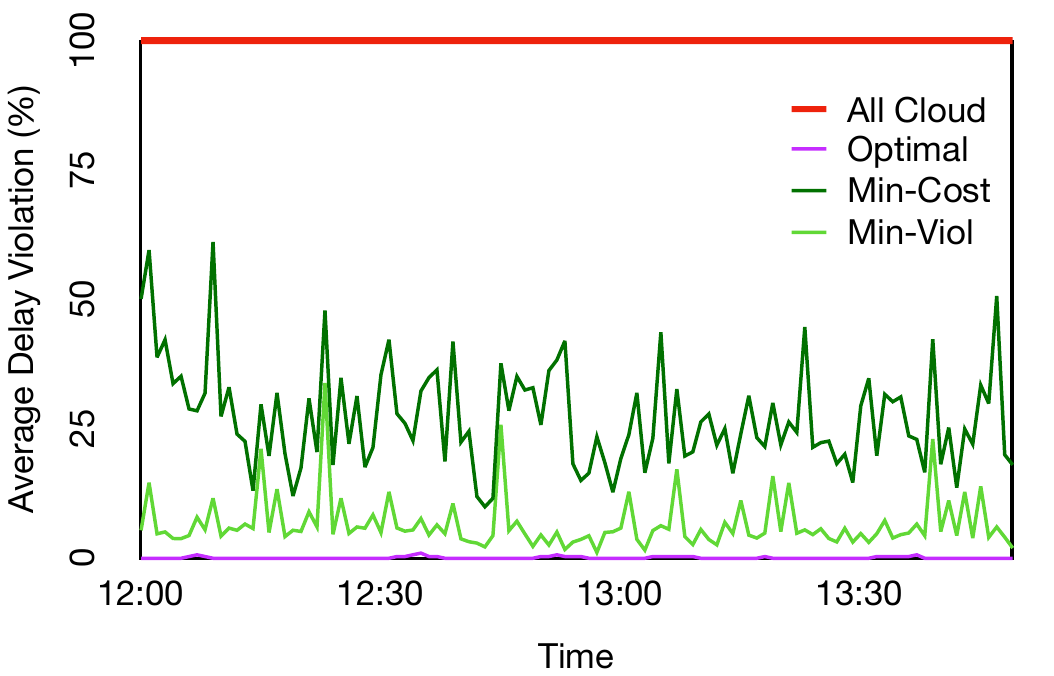}
        \label{trace2-viol}}\\
        \subfloat[{\scriptsize Average Number of Fog Services}]{\includegraphics[width=0.98\linewidth]{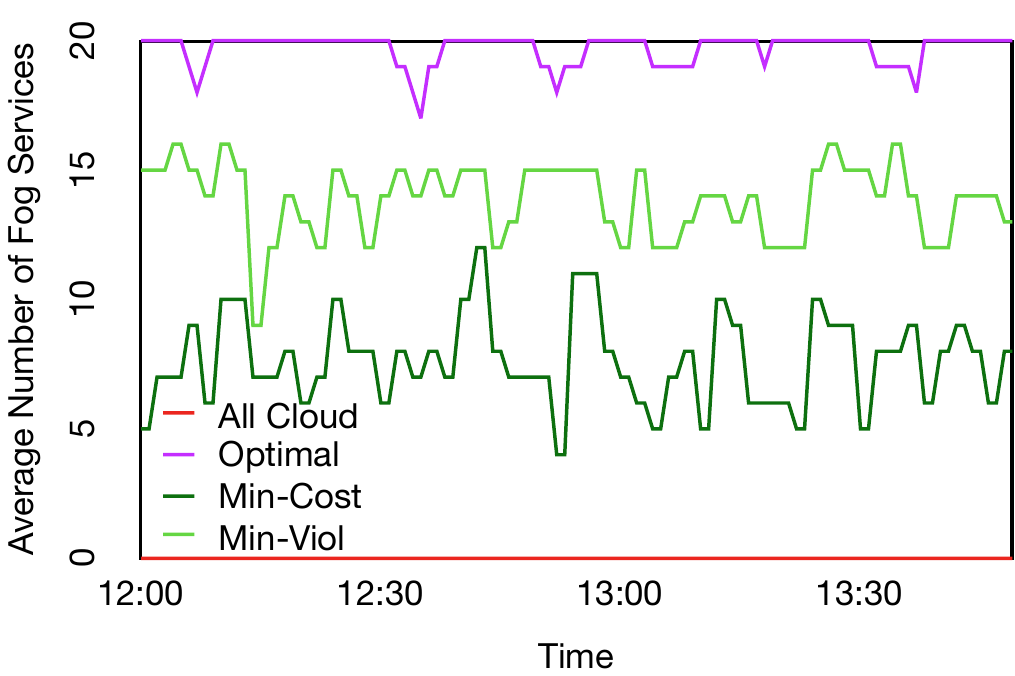}
        \label{trace2-services}}\\
        \subfloat[{\scriptsize Average Number of Cloud Services}]{\includegraphics[width=\linewidth]{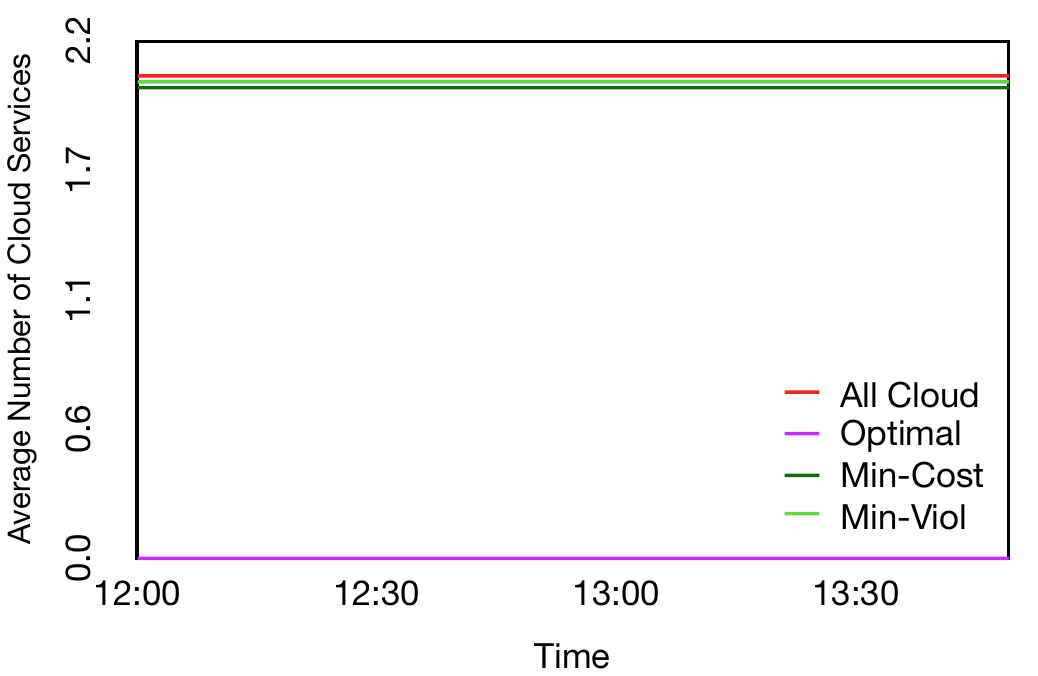}
        \label{trace2-services-cloud}}
        
\end{tabular}
\end{minipage} 
\caption{Simulation Results. Left: Experiment-1 (48-hour trace). Right: Experiment-2 (2-hour trace).} 
    \label{sim-results-1}
\end{figure}

This experiment is for studying the benefits of our proposed greedy algorithms under real-world traffic traces. The figures on the left column of Fig. \ref{sim-results-1} show the results of experiment-1 and are obtained using the 48 hours of trace data of {\em 2017/04/12-13}. In experiment-1, both the interval of traffic change and $\tau_a$ are 15 minutes. Fig. \ref{trace1} represents the normalized incoming traffic rates to the fog nodes in experiment-1. On the next row, Fig. \ref{trace1-delay} demonstrates the average service delay of the IoT nodes. Since there is no fog computing, All Cloud results in the highest service delay. Min-Viol achieves the lowest delay, as its goal is to minimize the violation. Min-Cost fluctuates around Static Fog, since, in a sense, Min-Cost is the dynamic variation of Static Fog. The reason Min-Cost has larger service delay than Static Fog will become clear soon.

On the third row, Fig. \ref{trace1-cost} shows the average cost of the various methods over time. All Cloud has the highest cost because it cannot satisfy the tight QoS requirement of the MAR applications and it incurs in a noticeable service penalty. Even though Min-Cost is envisioned for minimizing cost, it does not achieve the lowest cost. Interestingly, Min-Viol achieves the lowest cost, primarily as it minimizes the violation. The Static Fog's cost is more than that of Min-Cost, as its resulting static placement cannot keep up with the changing traffic demand.

Figure \ref{trace1-viol} illustrates the percentage of delay violations in the various methods. Clearly, All Cloud has the highest delay violation. Min-Viol's great performance is evident as it has the lowest delay violation of around 3\%. The delay violations of Static Fog and Min-Cost are comparable; when the magnitude of the traffic is small, Min-Cost deploys fewer fog services to reduce cost, thus incurs more delay violations. 

The figures on the last two rows of the left column of Fig. \ref{sim-results-1} show the average number of services deployed on the fog nodes (Fig. \ref{trace1-services}) and the cloud servers (Fig. \ref{trace1-services-cloud}). All Cloud does not deploy any services on fog nodes while it deploys all of the services in the cloud. As expected, the number of deployed services on the fog nodes for the Static Fog approach does not change over time. It is evident that Min-Viol deploys more services on the fog nodes than Min-Cost, to minimize the service delay, thus minimizing the delay violations. Among fog approaches, Min-Cost (roughly) has the lowest number of the deployed services on the fog nodes and the most number of the deployed services on the cloud server, as it tries to maintain a low-cost deployment. It is interesting to note the shape of the deployed services using Min-Cost; it seems that when the traffic rate is large, Min-Cost prefers deploying services in the cloud to deploying services on the fog nodes to minimize the cost.

\subsubsection{Experiment-2}
In experiment-2 we compare the performance of our two proposed greedy algorithms with that of the optimal solution achieved using the optimization problem introduced in Section \ref{QDFSP-INLP}. To solve the INLP problem, the Java program tries all the possible combinations of boolean variables $x_{aj}$ and $x'_{ak}$, and finds the placement that minimizes the cost. The figures in the right column of Fig. \ref{sim-results-1} show the results of this experiment and are obtained using 2 hours (12:00PM-2:00PM) of trace data of {\em 2017/04/12}. In experiment-2, the interval of traffic change is 1 minute and $\tau_a$ is set to 2 minutes.

Figure \ref{trace2} shows the normalized incoming traffic rates to the fog nodes. Figure \ref{trace2-delay} shows the average service delay, and we observe that All Cloud has the highest average service delay. Optimal and Min-Viol achieve the lowest average service delays, while Min-Cost's average service delay is larger. Figure \ref{trace2-cost} illustrates the average cost of the methods. As expected, All Cloud again has the highest cost, mainly due to its high delay violation, whereas Optimal brings the lowest cost since it finds the optimal solution to the cost minimization problem. Min-Viol has a slightly lower cost compared to Min-Cost, and its cost gets very close to that of the Optimal. 

Figure \ref{trace2-viol} illustrates the percentage of the delay violation in the various methods. All Cloud has the highest delay violation rate, whereas Min-Viol achieves the closest performance to that of the optimal service deployment. Finally, the last two rows on the right column of Fig. \ref{sim-results-1} show the average number of services deployed on the fog nodes (Fig. \ref{trace2-services}) and the cloud servers (Fig. \ref{trace2-services-cloud}). All Cloud deploys all the services in the cloud. Conversely, Optimal deploys all the services on the fog nodes. 

From this experiment, we can observe that Min-Viol achieves a closer performance to the optimal service deployment than the Min-Cost algorithm. The good performance of the Min-Viol algorithm is due to its aggressiveness on deploying more services on the fog nodes to minimize the violations. Moreover, since the application in this paper is assumed to have a high service delay penalty, Min-Viol's more deployed services on the fog nodes is the reason for its superior performance. 

\subsubsection{Experiment-3}
\begin{figure}[!t]
    \begin{minipage}{.48\linewidth}
    \begin{tabular}{l}
        \subfloat[{\scriptsize Service Delay vs. Threshold}]{\includegraphics[width=\linewidth]{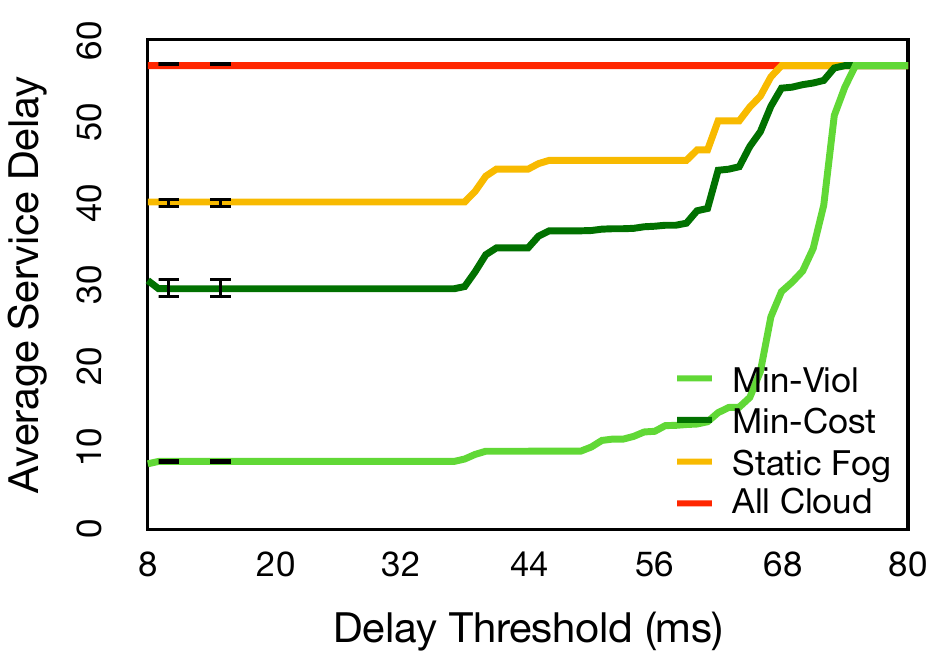}
        \label{thresh-delay}}\\
    \subfloat[{\scriptsize Cost vs. Threshold}]{\includegraphics[width=\linewidth]{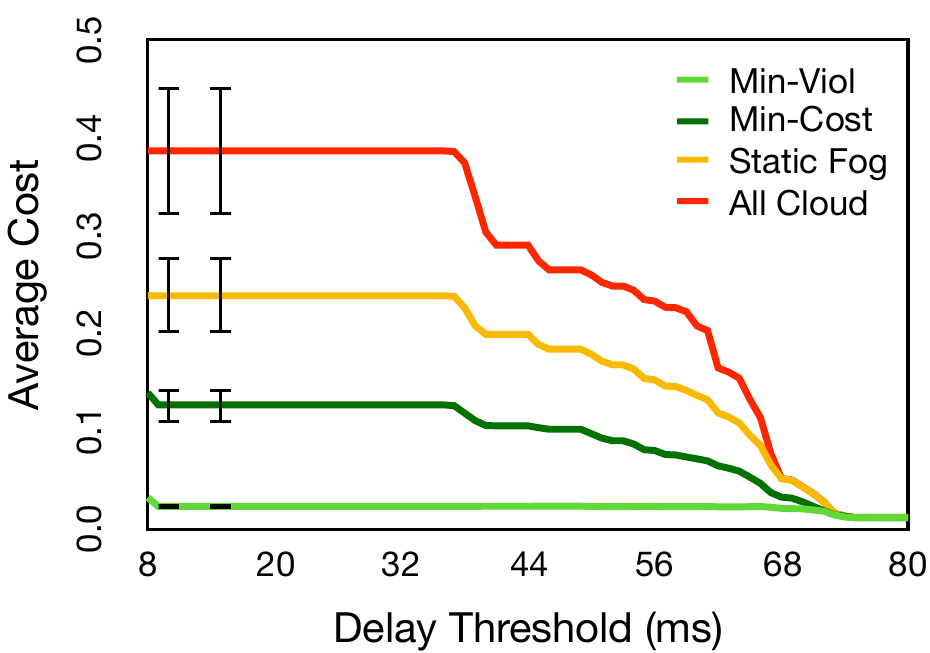}
        \label{thresh-cost}}\\
    \subfloat[{\scriptsize Delay Violation vs. Threshold}]{\includegraphics[width=\linewidth]{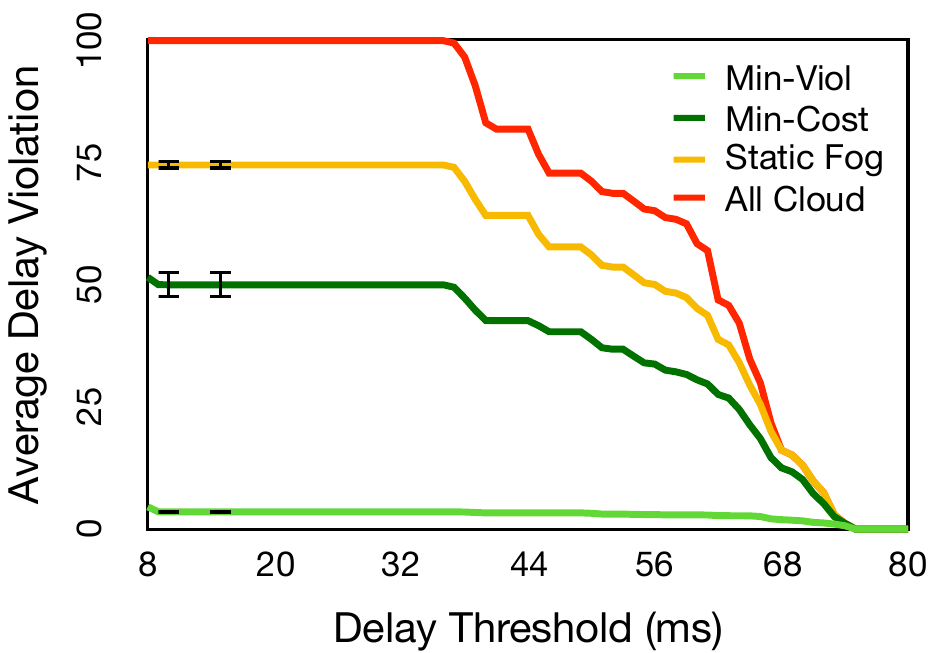}
        \label{thresh-viol}}\\
    \subfloat[{\scriptsize Fog Services vs. Threshold}]{\includegraphics[width=\linewidth]{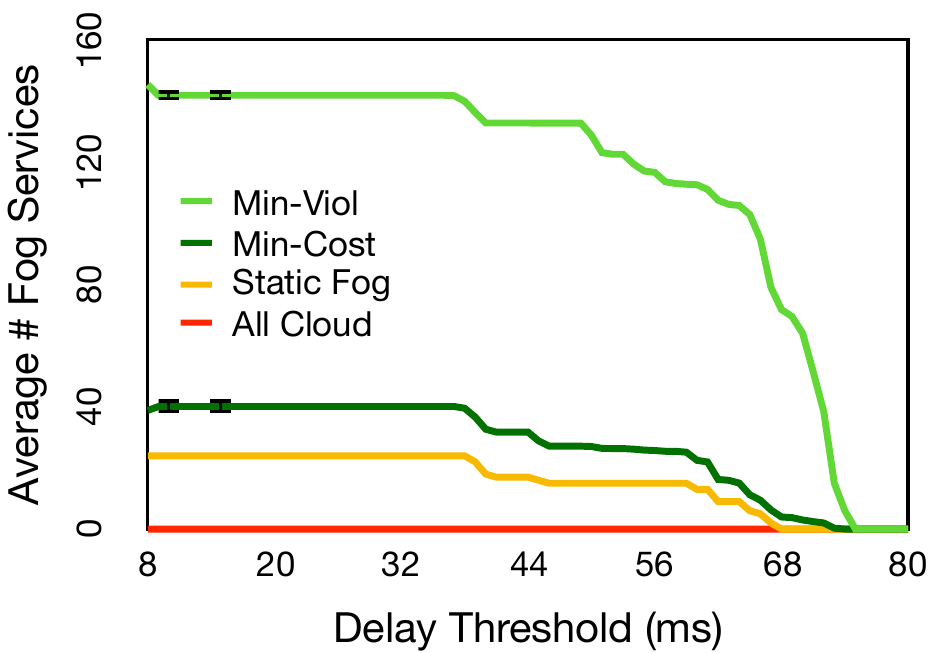}
        \label{thresh-services}}\\
        \subfloat[{\scriptsize Cloud Services vs. Threshold}]{\includegraphics[width=\linewidth]{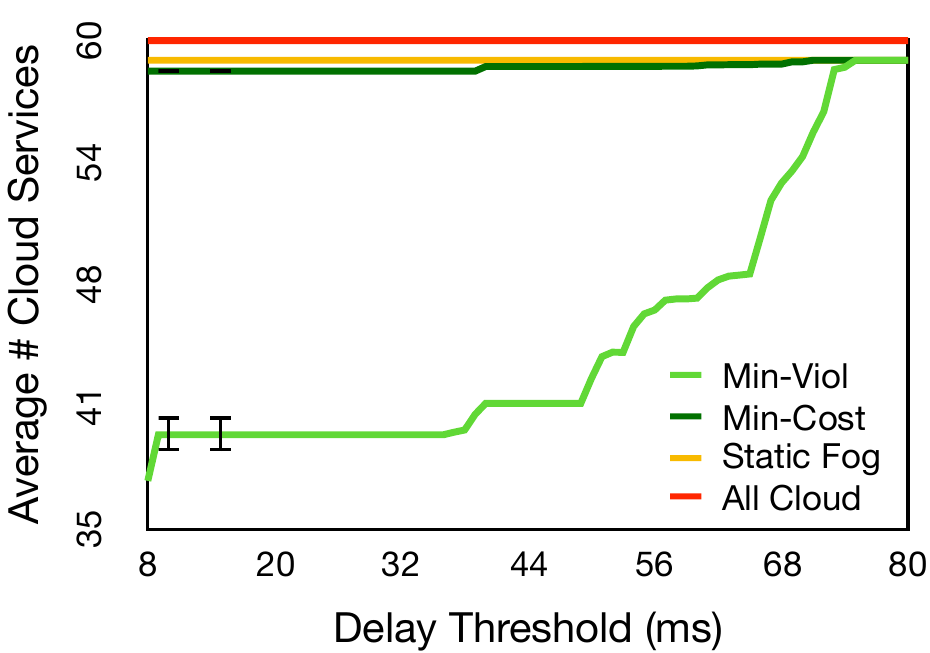}
        \label{thresh-services-cloud}}
    \end{tabular}
    \end{minipage}%
    ~
    \begin{minipage}{.48\linewidth}
    \begin{tabular}{l}
    \subfloat[{\scriptsize Service Delay vs. Reconfig. Interval}]{\includegraphics[width=\linewidth]{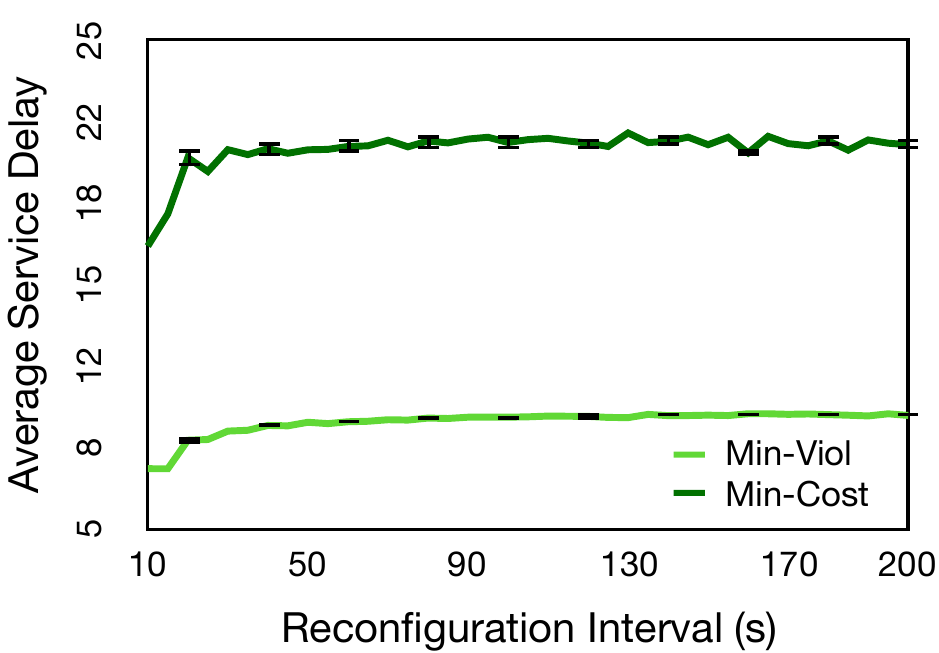}
        \label{tau-delay}}\\
    \subfloat[{\scriptsize Cost vs. Reconfig. Interval}]{\includegraphics[width=\linewidth]{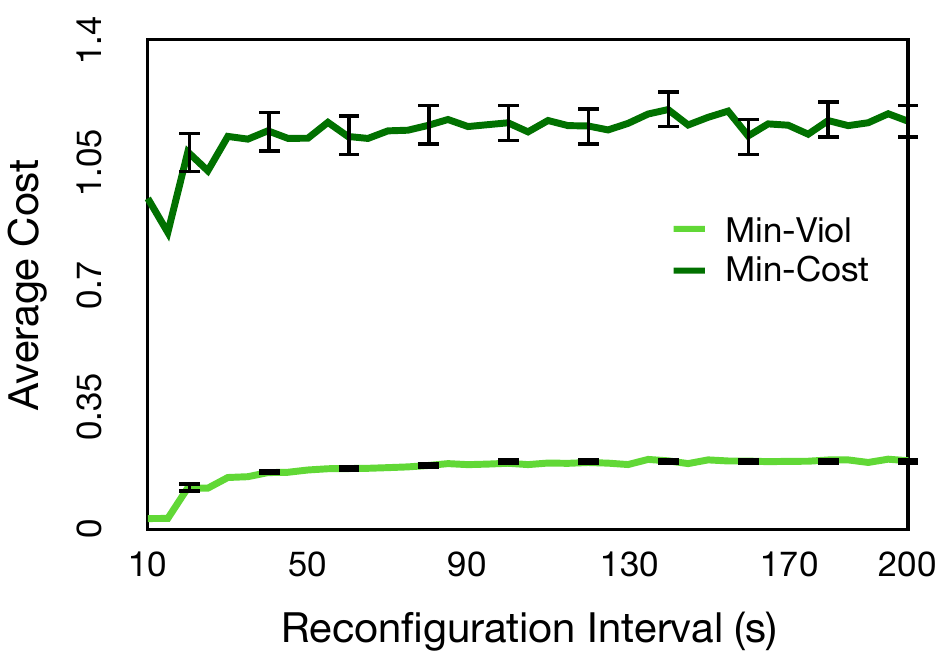}
        \label{tau-cost}}\\
    \subfloat[{\scriptsize Delay Violation vs. Reconfig. Interval}]{\includegraphics[width=\linewidth]{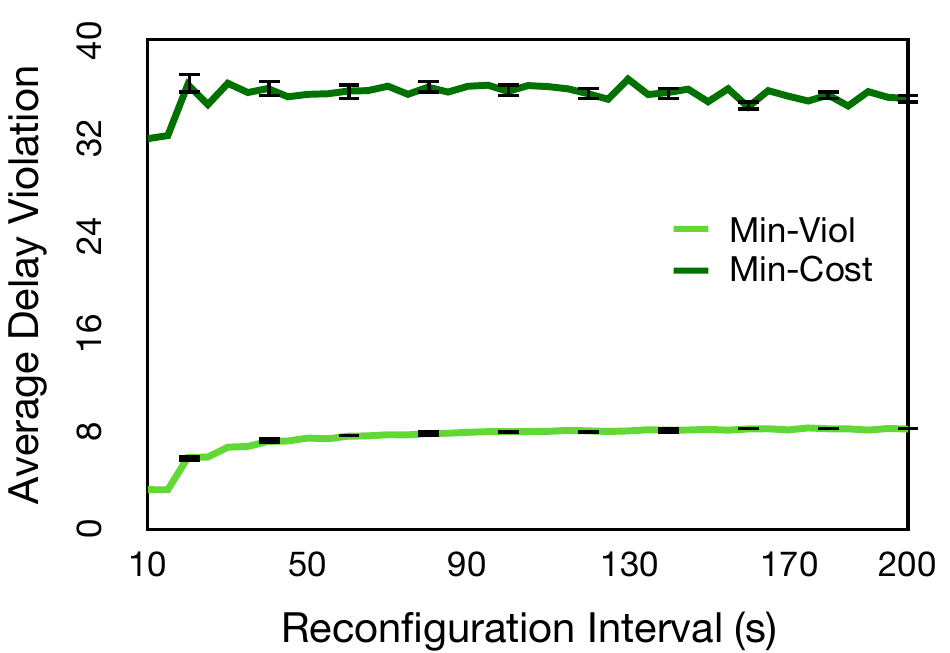}
        \label{tau-viol}}\\
    \subfloat[{\scriptsize Fog Services vs. Reconfig. Interval}]{\includegraphics[width=\linewidth]{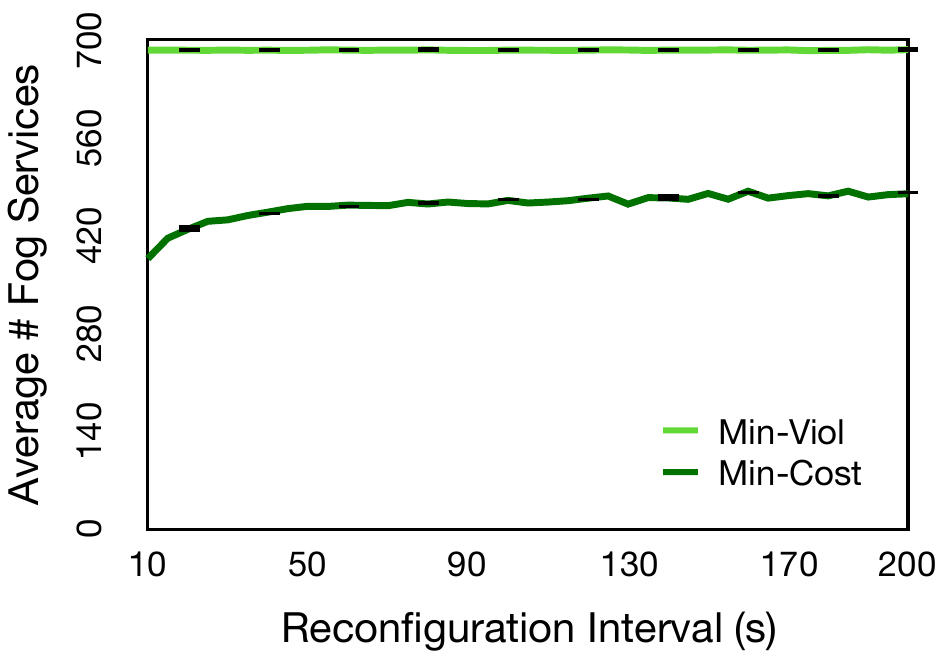}
        \label{tau-services}}\\
    \subfloat[{\scriptsize Cloud Services vs. Reconfig. Interval}]{\includegraphics[width=\linewidth]{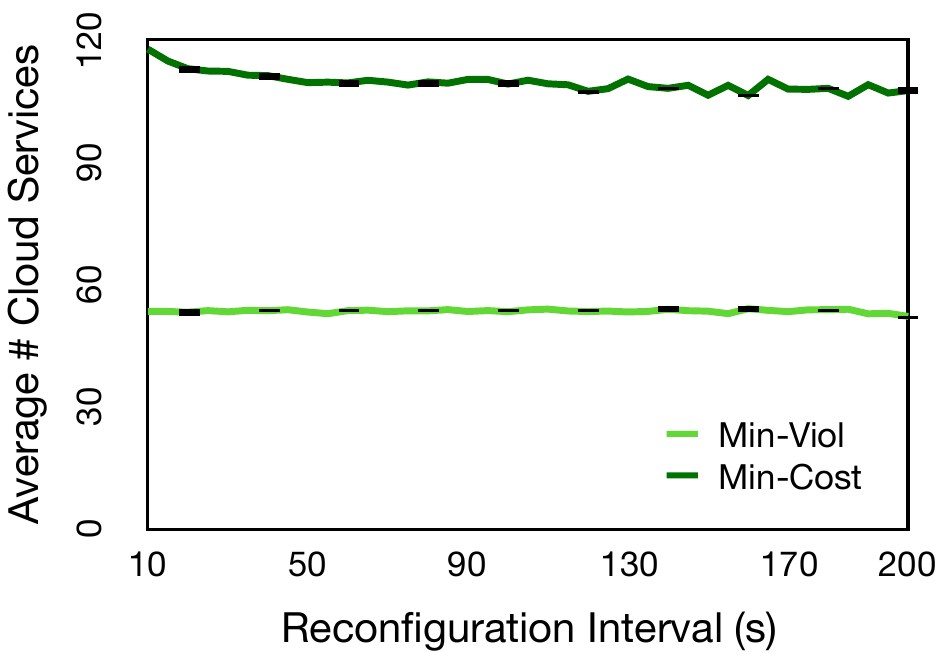}
        \label{tau-services-cloud}}
\end{tabular}
\end{minipage}

   \caption{Simulation Results. Left: Experiment-3 (Impact of delay threshold $th_a$). Right: Experiment-4 (Impact of reconfiguration interval length $\tau_a$). \label{sim-results-2}
}
    \end{figure}

The next set of diagrams shown in the left column of Fig. \ref{sim-results-2} illustrate the impact of delay threshold for service $a$ ($th_a$) on our proposed algorithms. These figures are obtained using the trace of 4 hours (12:00PM-4:00PM) of {\em 2017/04/12}. The error bars on this set of figures show the 99\% confidence intervals of the results. When the width of the confidence interval, no error bar is shown. The interval of the traffic change and $\tau_a$ are 10 seconds. The rest of the parameters of the simulation are the same as before. 

Figure \ref{thresh-delay} depicts how delay threshold affects the average service delay. When the delay threshold increases, the average service delay of all the fog approaches (Min-Viol, Min-Cost, and Static Fog) increases, and finally reaches to that of All Cloud. This is because when the delay threshold is large, the greedy algorithms tend to deploy fewer services on the fog nodes; hence requests will have large service delays as they will be served in the cloud. As expected, Min-Viol has the lowest average service delay among all the approaches. Min-Cost achieves the second lowest average service delay. 

Figure \ref{thresh-cost} illustrates the decrease in the average cost of all four approaches when the delay threshold increases. This is the case because when the delay threshold is high, fewer services are deployed on the fog nodes, and fewer requests violate the delay threshold requirements. When the delay threshold is high, there will be fewer delay violations, and the cost of All Cloud gets closer to that of the other fog approaches. When the delay threshold is larger than 75 ms, there will be no delay violation even if the services are not deployed on the fog nodes. Hence, when the delay threshold is larger than 75 ms, all the services will be deployed in the cloud, and all the approaches will have the same cost value. When the delay threshold is not too large, Min-Cost and Min-Viol have the lowest cost. 

Figure \ref{thresh-viol} shows the performance of the four approaches with regards to delay violations. Min-Viol's delay violation is the lowest, while the All Cloud approach has the highest delay violation. The violations of Min-Cost and Static Fog are moderate because they are not tuned for the goal of minimizing delay violation. As the delay threshold becomes higher, the delay violation of all the approaches becomes smaller. The delay violation of all the approaches does not change when the delay threshold is less than 38 ms. The reason is, with the values of the simulation parameters in the current setup, when the delay threshold is below 38 ms, all the approaches place the same number of services on the fog nodes and the cloud servers. Nevertheless, when the delay threshold becomes greater than 38 ms, the average delay violation becomes lower. This period of constant delay violation is also evident in Fig. \ref{thresh-services} and Fig. \ref{thresh-services-cloud}. In both figures, the number of deployed services on fog nodes and cloud servers remains constant when the delay threshold is less than 38 ms. As the delay threshold becomes higher, fewer services are deployed on the fog nodes, and a greater number of services are deployed on the cloud servers.

\subsubsection{Experiment-4}
The next set of diagrams shown in the right column of Fig. \ref{sim-results-2} depict the impact of the reconfiguration interval length ($\tau_a$) on our proposed algorithms. We use the DTMC-based traffic trace generator for this experiment for the traffic trace. Similar to experiment-3, the error bars on this set of figures show the 99\% confidence intervals of the results. Min-Viol and Min-Cost are shown on these figures since the reconfiguration interval is not relevant for All Cloud and Static Fog. The interval of the traffic change is 10 seconds and the length of the reconfiguration interval varies from 10 to 200 seconds. 

Figure \ref{tau-delay} illustrates how the length of the reconfiguration interval affects the average service delay. We can see that when $\tau_a$ becomes larger, the average service delay becomes larger as well, simply because the greedy algorithms are not run frequently. Similarly, Fig. \ref{tau-cost} shows how the average cost of both algorithms increases as the length of the reconfiguration interval becomes greater. This is again due to the infrequent run of the algorithms that incurs more service penalty. 

The results in Fig. \ref{tau-viol} also show the increase in the average delay violations as $\tau_a$ becomes greater. Finally, the last two diagrams in the right column of Fig. \ref{sim-results-2} show the number of deployed services on the fog nodes and cloud servers as a function of $\tau_a$. Min-Viol maintains the same number of deployed services, whereas the length of the reconfiguration interval affects the number of deployed services in Min-Cost. 

We reason that there exists a fundamental tradeoff for choosing the appropriate length of the reconfiguration interval: the length of the reconfiguration interval must be small enough so that the frequent running of the greedy algorithms can reduce the average service delay, violation, and finally can minimize the cost. On the other hand, the reconfiguration interval must be long enough for the greedy algorithms to finish running. The length of the reconfiguration interval must be chosen according to the aforementioned trade-off. In the next experiment, we numerically measure the actual runtime of the greedy algorithms in different settings. 

\subsubsection{Experiment-5}
In this experiment, we analyze the scalability of our proposed greedy algorithms using larger network topologies and more services. In Section \ref{complexity} we discussed the asymptotic complexity of Min-Cost and Min-Viol. In this section, we discuss our measurements of the actual running time of these algorithms. Since the number of regions/cities representing fog nodes is limited in the real-world MAWI traffic traces, we use the DTMC-based traffic trace generator for this experiment. 

The results are shown in Table \ref{scalability}. The numbers in the columns titled {\em Time} in Table \ref{scalability} are the averages of running the greedy algorithms for the corresponding number of services 10 times. For instance, the numbers in the time column of the first and fourth rows of Table \ref{scalability} represent the average running times for 10 runs of the greedy algorithms over 100 services. The running time of the algorithms is precisely measured using the {\em time} command in Unix. The initialization time is subtracted from the total time to get the exact running time of the algorithms. The computer we used for this experiment has the following specifications: 4-Core Intel Xeon E5620, 12GB RAM, CentOS 6.9 (Kernel 2.6.32-696), JRE 1.8.0$\_$171, JVM 25.171.

Recall that the asymptotic complexity of both Min-Cost and Min-Viol is $O(|A||F|^2)$. This is also evident by the results in the Table \ref{scalability}; the algorithms scale linearly with the number of services (first 3 rows) and the running times stay below 1 second. However, the algorithms scale quadratically with the number of fog nodes (last 3 rows). For instance, when we have 10,000 fog nodes, Min-Cost takes about 3 seconds and Min-Viol takes around 2 minutes to finish the service deployment of one service. The total time for running the Min-Cost and Min-Viol for all services, in this case, would be around 5 minutes and 3 hours, respectively. 

\begin{table}[!t]
\caption{\label{scalability}Scalability of Greedy Methods. Number of cloud servers is 3. Time is an average for individual service. $q_a=90\%$}
\newcolumntype{L}[1]{>{\raggedright\let\newline\\\arraybackslash\hspace{0pt}}m{#1}}
    \centering
    {\footnotesize
        {\renewcommand\arraystretch{1.0} 
            \begin{tabular}{|L{0.17\columnwidth}|L{0.15\columnwidth}|L{0.24\columnwidth}|L{0.24\columnwidth}|}
                \hline
                \# Fog Nodes & \# Services & Time (Min-Cost) & Time (Min-Viol)\\
                \hline\hline
                100 & 100 & 8ms & 6ms\\
                100 & 1000 & 11ms & 13ms\\
                    100 & 10000 & 60ms & 272ms\\
                    \hline
                    100 & 100 & 8ms & 6ms\\
                    1000 & 100 & 384ms & 303ms\\
                    10000 & 100 & 3s 199ms & 1m 58s 992ms\\                
                \hline
            \end{tabular}
        }}
        
    \end{table}
    
\subsection{Discussion}
It can be seen that Min-Cost has generally lower running time (is faster) in bigger settings since its main loop conditions do not depend on the value of $q_a$ (they depend on cost). Conversely, since the main loops of Min-Viol depend on the value of $q_a$, Min-Viol takes longer to finish when the quality of service value is strict (e.g. $q_a=99.9\%$). Thus, in general, Min-Viol is slower than Min-Cost. Nevertheless, as seen in the previous experiments, Min-Viol has lower average service delay and lower delay violations than Min-Cost. It is clear now that Min-Viol's superior performance is at the cost of its slower runtime. The reconfiguration interval length must be chosen big enough so that the chosen greedy algorithm can finish deploying services during this interval. 

As discussed before, to address the \qdfsp~problem, solving the optimization problem periodically may be not feasible, especially for large networks. The proposed greedy algorithms can be called periodically as alternative approaches for addressing the \qdfsp~problem. The results show the performance of Min-Cost and Min-Viol algorithms compared to the optimal. 

Nevertheless, it may not be also efficient to run the greedy algorithms too frequently, as it is not always good to deploy (or release) services in response to a short-lived traffic peak (or drop). We refer to the issue of frequent deploy and release as {\em impatient provisioning}, and it can happen when the traffic rate fluctuates rapidly. The limitation of the current \schemeName~framework is that it lacks a standard mechanism to handle impatient provisioning; one has to manually choose an appropriate length for the reconfiguration interval $\tau_a$, considering the discussed trade-off in experiment-4 and the impatient provisioning issue. 

One way to address impatient provisioning is to choose a greater value for reconfiguration interval length and consider the average of the traffic in those durations as the input traffic to the algorithms. Another method would be using online learning algorithms, such as {\em follow the regularized leader (FTRL)} \cite{mcmahan2011follow} for traffic prediction. If the traffic is predicted beforehand, peaks and drops could be predicted and the decision to run the greedy algorithms can be made more intelligently. To improve \schemeName, we plan to incorporate learning methods for traffic prediction in our future work. 
\section{Conclusion} \label{Conclusion}
Fog is a continuum filling the gap between cloud and IoT, which brings low latency, location awareness, and reduced bandwidth to the IoT applications. We discussed how \schemeName~could benefit Fog Service Providers and their clients, in terms of improved QoS and cost savings. For addressing the \qdfsp~problem, a possible INLP formulation and two greedy algorithms are introduced that must be run periodically. Finally, we presented the results of our experiments based on real-world traffic traces and a DTMC-based traffic generator. Both Min-Viol and Min-Cost have the same asymptotic complexity, however, Min-Cost is faster than Min-Viol, especially when there are more fog nodes and services. Except for the optimal deployment (achieved by solving the INLP), Min-Viol has the lowest average service delay and average delay violations. We saw that Min-Viol's superior performance comes at the cost of a slower runtime. We also discussed the fundamental trade-off for choosing the appropriate length of the reconfiguration interval. 

As future work, it is interesting to see how the distance of the FSC, relative to the fog nodes, can affect the proposed scheme. Moreover, \qdfsp~is achieved through either deploying new service(s) or releasing existing service(s) from fog nodes, which means the decision variables are binary. However, considering non-binary variables (that is considering scaling up and down service or container capacities) as a response to load variation may be a future research direction. Lastly, as briefly discussed in the paper, additional future directions are to (i) consider stateful fog services and related migration issues and (ii) design a protocol for fog service discovery (iii) incorporate learning methods for traffic prediction. 





%
\bibliography{Master}{}

\begin{thebibliography}{10}

\bibitem{byers2017architectural}
C.~C. Byers, ``Architectural imperatives for fog computing: Use cases,
  requirements, and architectural techniques for fog-enabled iot networks,''
  {\em IEEE Communications Magazine}, vol.~55, no.~8, pp.~14--20, 2017.

\bibitem{extendingCloudtoEdge}
R.~S. Montero, E.~Rojas, A.~A. Carrillo, and I.~M. Llorente, ``Extending the
  cloud to the network edge.,'' {\em IEEE Computer}, vol.~50, no.~4,
  pp.~91--95, 2017.

\bibitem{fog-survey}
C.~Mouradian, D.~Naboulsi, S.~Yangui, R.~H. Glitho, M.~J. Morrow, and P.~A.
  Polakos, ``A comprehensive survey on fog computing: State-of-the-art and
  research challenges,'' {\em IEEE Communications Surveys \& Tutorials},
  vol.~20, no.~1, pp.~416--464, 2017.

\bibitem{edge-survey}
C.~Li, Y.~Xue, J.~Wang, W.~Zhang, and T.~Li, ``Edge-oriented computing
  paradigms: A survey on architecture design and system management,'' {\em ACM
  Computing Surveys (CSUR)}, vol.~51, no.~2, p.~39, 2018.

\bibitem{taleb2017multi}
T.~Taleb, K.~Samdanis, B.~Mada, H.~Flinck, S.~Dutta, and D.~Sabella, ``On
  multi-access edge computing: A survey of the emerging 5g network edge cloud
  architecture and orchestration,'' {\em IEEE Communications Surveys \&
  Tutorials}, vol.~19, no.~3, pp.~1657--1681, 2017.

\bibitem{openfog}
OpenFogConsortium, ``Openfog reference architecture for fog computing,'' 2017.
\newblock [Online]. Available: https://www.openfogconsortium.org/ra/, February
  2017.

\bibitem{NIST}
M.~Iorga, L.~Feldman, R.~Barton, M.~J. Martin, N.~S. Goren, and C.~Mahmoudi,
  ``Fog computing conceptual model,'' 2018.
\newblock NIST Special Publication 500-325.

\bibitem{saurez2016incremental}
E.~Saurez, K.~Hong, D.~Lillethun, U.~Ramachandran, and B.~Ottenw{\"a}lder,
  ``Incremental deployment and migration of geo-distributed situation awareness
  applications in the fog,'' in {\em 10th ACM International Conference on
  Distributed and Event-based Systems}, pp.~258--269, 2016.

\bibitem{reviewer-asked-us-to-cite-his-work}
K.~C. Okafor, I.~E. Achumba, G.~A. Chukwudebe, and G.~C. Ononiwu, ``Leveraging
  fog computing for scalable iot datacenter using spine-leaf network
  topology,'' {\em Journal of Electrical and Computer Engineering}, vol.~2017,
  2017.

\bibitem{azure}
``Azure iot edge.''
\newblock https://azure.microsoft.com/services/iot-edge.

\bibitem{gc}
``Google cloud iot edge.''
\newblock https://cloud.google.com/iot-edge.

\bibitem{aws}
``Amazon aws greengrass.''
\newblock https://aws.amazon.com/greengrass.

\bibitem{abdelwahab2016flocking}
S.~Abdelwahab and B.~Hamdaoui, ``Flocking virtual machines in quest for
  responsive iot cloud services,'' in {\em Communications (ICC), 2017 IEEE
  International Conference on}, pp.~1--6, IEEE, 2017.

\bibitem{zhang2016segue}
W.~Zhang, Y.~Hu, Y.~Zhang, and D.~Raychaudhuri, ``Segue: Quality of service
  aware edge cloud service migration,'' in {\em Cloud Computing Technology and
  Science (CloudCom), 2016 IEEE International Conference on}, pp.~344--351,
  2016.

\bibitem{ma2018efficient}
L.~Ma, S.~Yi, N.~Carter, and Q.~Li, ``Efficient live migration of edge services
  leveraging container layered storage,'' {\em IEEE Transactions on Mobile
  Computing}, 2018.

\bibitem{clark2005live}
C.~Clark, K.~Fraser, S.~Hand, J.~G. Hansen, E.~Jul, C.~Limpach, I.~Pratt, and
  A.~Warfield, ``Live migration of virtual machines,'' NSDI'05, pp.~273--286,
  USENIX Association, 2005.

\bibitem{thesis-migration}
F.~Zhang, {\em Challenges and New Solutions for Live Migration of Virtual
  Machines in Cloud Computing Environments}.
\newblock PhD thesis, Georg-August-Universit{\"a}t G{\"o}ttingen, 2018.

\bibitem{ha2017you}
K.~Ha, Y.~Abe, T.~Eiszler, Z.~Chen, W.~Hu, B.~Amos, R.~Upadhyaya, P.~Pillai,
  and M.~Satyanarayanan, ``You can teach elephants to dance: agile vm handoff
  for edge computing,'' in {\em Proceedings of the Second ACM/IEEE Symposium on
  Edge Computing}, p.~12, ACM, 2017.

\bibitem{service-migration}
A.~Machen, S.~Wang, K.~K. Leung, B.~J. Ko, and T.~Salonidis, ``Live service
  migration in mobile edge clouds,'' {\em IEEE Wireless Communications},
  vol.~25, no.~1, pp.~140--147, 2018.

\bibitem{chaufournier2017fast}
L.~Chaufournier, P.~Sharma, F.~Le, E.~Nahum, P.~Shenoy, and D.~Towsley, ``Fast
  transparent virtual machine migration in distributed edge clouds,'' in {\em
  Proceedings of the Second ACM/IEEE Symposium on Edge Computing}, p.~10, ACM,
  2017.

\bibitem{oci}
M.~K{\"o}rner, T.~M. Runge, A.~Panda, S.~Ratnasamy, and S.~Shenker, ``Open
  carrier interface: An open source edge computing framework,'' in {\em
  Proceedings of the 2018 Workshop on Networking for Emerging Applications and
  Technologies}, pp.~27--32, ACM, 2018.

\bibitem{platform}
S.~Yangui, P.~Ravindran, O.~Bibani, R.~H. Glitho, N.~B. Hadj-Alouane, M.~J.
  Morrow, and P.~A. Polakos, ``A platform as-a-service for hybrid cloud/fog
  environments,'' in {\em Local and Metropolitan Area Networks (LANMAN), 2016
  IEEE International Symposium on}, pp.~1--7, 2016.

\bibitem{Indiefog}
C.~Chang, S.~N. Srirama, and R.~Buyya, ``Indie fog: An efficient fog-computing
  infrastructure for the internet of things,'' {\em Computer}, vol.~50, no.~9,
  pp.~92--98, 2017.

\bibitem{wang2017enorm}
N.~Wang, B.~Varghese, M.~Matthaiou, and D.~S. Nikolopoulos, ``Enorm: A
  framework for edge node resource management,'' {\em IEEE Transactions on
  Services Computing}, 2017.

\bibitem{incremental}
E.~Saurez, K.~Hong, D.~Lillethun, U.~Ramachandran, and B.~Ottenw{\"a}lder,
  ``Incremental deployment and migration of geo-distributed situation awareness
  applications in the fog,'' in {\em Proceedings of the 10th ACM International
  Conference on Distributed and Event-based Systems}, pp.~258--269, 2016.

\bibitem{picasso}
A.~Lertsinsrubtavee, M.~Selimi, A.~Sathiaseelan, L.~Cerd{\`a}-Alabern,
  L.~Navarro, and J.~Crowcroft, ``Information-centric multi-access edge
  computing platform for community mesh networks,'' in {\em Proceedings of the
  1st ACM SIGCAS Conference on Computing and Sustainable Societies},
  pp.~19:1--19:12, ACM, 2018.

\bibitem{zenith}
J.~Xu, B.~Palanisamy, H.~Ludwig, and Q.~Wang, ``Zenith: Utility-aware resource
  allocation for edge computing,'' in {\em Edge Computing (EDGE), 2017 IEEE
  International Conference on}, pp.~47--54, 2017.

\bibitem{zhang2017towards}
W.~Zhang, J.~Chen, Y.~Zhang, and D.~Raychaudhuri, ``Towards efficient edge
  cloud augmentation for virtual reality mmogs,'' in {\em Proceedings of the
  Second ACM/IEEE Symposium on Edge Computing}, p.~8, ACM, 2017.

\bibitem{migcep}
B.~Ottenw{\"a}lder, B.~Koldehofe, K.~Rothermel, and U.~Ramachandran, ``Migcep:
  operator migration for mobility driven distributed complex event
  processing,'' in {\em Proceedings of the 7th ACM international conference on
  Distributed event-based systems}, pp.~183--194, ACM, 2013.

\bibitem{dynamic-module-deploy}
H.-J. Hong, P.-H. Tsai, and C.-H. Hsu, ``Dynamic module deployment in a fog
  computing platform,'' in {\em Network Operations and Management Symposium
  (APNOMS), 2016 18th Asia-Pacific}, pp.~1--6, 2016.

\bibitem{asymptotically}
I.~Hou, T.~Zhao, S.~Wang, K.~Chan, {\em et~al.}, ``Asymptotically optimal
  algorithm for online reconfiguration of edge-clouds,'' in {\em Proceedings of
  the 17th ACM International Symposium on Mobile Ad Hoc Networking and
  Computing}, pp.~291--300, ACM, 2016.

\bibitem{placement1}
L.~Yang, J.~Cao, G.~Liang, and X.~Han, ``Cost aware service placement and load
  dispatching in mobile cloud systems,'' {\em IEEE Transactions on Computers},
  vol.~65, no.~5, pp.~1440--1452, 2016.

\bibitem{placement2}
S.~Wang, R.~Urgaonkar, T.~He, K.~Chan, M.~Zafer, and K.~K. Leung, ``Dynamic
  service placement for mobile micro-clouds with predicted future costs,'' {\em
  IEEE Transactions on Parallel and Distributed Systems}, vol.~28, no.~4,
  pp.~1002--1016, 2017.

\bibitem{placement3}
Q.~Zhang, Q.~Zhu, M.~F. Zhani, R.~Boutaba, and J.~L. Hellerstein, ``Dynamic
  service placement in geographically distributed clouds,'' {\em IEEE Journal
  on Selected Areas in Communications}, vol.~31, no.~12, pp.~762--772, 2013.

\bibitem{vne1}
M.~Yu, Y.~Yi, J.~Rexford, and M.~Chiang, ``Rethinking virtual network
  embedding: substrate support for path splitting and migration,'' {\em ACM
  SIGCOMM Computer Communication Review}, vol.~38, no.~2, pp.~17--29, 2008.

\bibitem{skarlat2017towards}
O.~Skarlat, M.~Nardelli, S.~Schulte, M.~Borkowski, and P.~Leitner, ``Optimized
  iot service placement in the fog,'' {\em Service Oriented Computing and
  Applications}, vol.~11, no.~4, pp.~427--443, 2017.

\bibitem{yigitoglu2017foggy}
E.~Yigitoglu, M.~Mohamed, L.~Liu, and H.~Ludwig, ``Foggy: a framework for
  continuous automated iot application deployment in fog computing,'' in {\em
  2017 IEEE International Conference on AI \& Mobile Services (AIMS)},
  pp.~38--45, IEEE, 2017.

\bibitem{mist-cisco}
A.~Davies, ``Cisco pushes iot analytics to the extreme edge with mist
  computing.''
\newblock [Online]. Available:
  http://rethinkresearch.biz/articles/cisco-pushes-iot-analytics-extreme-edge-mist-computing-2,
  Blog, Rethink Research.

\bibitem{NetFATE}
G.~Faraci and G.~Schembra, ``An analytical model to design and manage a green
  sdn/nfv cpe node,'' {\em IEEE Transactions on Network and Service
  Management}, vol.~12, no.~3, pp.~435--450, 2015.

\bibitem{faraci2015analytical}
G.~Faraci and G.~Schembra, ``An analytical model to design and manage a green
  sdn/nfv cpe node,'' {\em IEEE Transactions on Network and Service
  Management}, vol.~12, no.~3, pp.~435--450, 2015.

\bibitem{ashkan-fog-delay}
A.~Yousefpour, G.~Ishigaki, R.~Gour, and J.~P. Jue, ``On reducing iot service
  delay via fog offloading,'' {\em IEEE Internet of Things Journal}, vol.~5,
  pp.~998--1010, April 2018.

\bibitem{edge-resource}
C.~Anglano, M.~Canonico, and M.~Guazzone, ``Profit-aware resource management
  for edge computing systems,'' in {\em Proceedings of the 1st International
  Workshop on Edge Systems, Analytics and Networking}, pp.~25--30, ACM, 2018.

\bibitem{ERO}
D.~Kim, H.~Lee, H.~Song, N.~Choi, and Y.~Yi, ``On the economics of fog
  computing: Inter-play among infrastructure and service providers, users, and
  edge resource owners,'' in {\em 2018 IEEE International Conference on
  Communications (ICC)}, pp.~1--6, IEEE, 2018.

\bibitem{sdn-monitor}
J.~Su{\'a}rez-Varela and P.~Barlet-Ros, ``Sbar: Sdn flow-based monitoring and
  application recognition,'' in {\em Proceedings of the Symposium on SDN
  Research}, p.~22, ACM, 2018.

\bibitem{sdn-monitor2}
D.-H. Luong, A.~Outtagarts, and A.~Hebbar, ``Traffic monitoring in software
  defined networks using opendaylight controller,'' in {\em International
  Conference on Mobile, Secure and Programmable Networking}, pp.~38--48,
  Springer, 2016.

\bibitem{container-registry}
https://docs.microsoft.com/azure/iot-edge/tutorial-deploy-function.

\bibitem{Cloudpath}
S.~H. Mortazavi, M.~Salehe, C.~S. Gomes, C.~Phillips, and E.~de~Lara,
  ``Cloudpath: a multi-tier cloud computing framework,'' in {\em Proceedings of
  the Second ACM/IEEE Symposium on Edge Computing}, p.~20, ACM, 2017.

\bibitem{container-as-service}
K.~Kaur, T.~Dhand, N.~Kumar, and S.~Zeadally, ``Container-as-a-service at the
  edge: Trade-off between energy efficiency and service availability at fog
  nano data centers,'' {\em IEEE Wireless Communications}, vol.~24, no.~3,
  pp.~48--56, 2017.

\bibitem{m-m-c-1}
M.~Jia, J.~Cao, and W.~Liang, ``Optimal cloudlet placement and user to cloudlet
  allocation in wireless metropolitan area networks,'' {\em IEEE Transactions
  on Cloud Computing}, 2015.

\bibitem{m-m-c-2}
Z.~Zhou, J.~Feng, L.~Tan, Y.~He, and J.~Gong, ``An air-ground integration
  approach for mobile edge computing in iot,'' {\em IEEE Communications
  Magazine}, vol.~56, no.~8, pp.~40--47, 2018.

\bibitem{m-m-c-3}
L.~Liu, X.~Guo, Z.~Chang, and T.~Ristaniemi, ``Joint optimization of energy and
  delay for computation offloading in cloudlet-assisted mobile cloud
  computing,'' {\em Wireless Networks}, pp.~1--14, July 2018.

\bibitem{bolch2006queueing}
G.~Bolch, S.~Greiner, H.~De~Meer, and K.~S. Trivedi, {\em Queueing networks and
  Markov chains: modeling and performance evaluation with computer science
  applications}.
\newblock John Wiley \& Sons, 2006.

\bibitem{wide}
``Wide mawi working group traffic archive,'' http://mawi.wide.ad.jp.

\bibitem{iot-survey}
A.~Al-Fuqaha, M.~Guizani, M.~Mohammadi, M.~Aledhari, and M.~Ayyash, ``Internet
  of things: A survey on enabling technologies, protocols, and applications,''
  {\em IEEE Communications Surveys \& Tutorials}, vol.~17, no.~4,
  pp.~2347--2376, 2015.

\bibitem{cooperative}
A.~Kapsalis, P.~Kasnesis, I.~S. Venieris, D.~I. Kaklamani, and C.~Z.
  Patrikakis, ``A cooperative fog approach for effective workload balancing,''
  {\em IEEE Cloud Computing}, vol.~4, no.~2, pp.~36--45, 2017.

\bibitem{impact}
K.~Ha, P.~Pillai, G.~Lewis, S.~Simanta, S.~Clinch, N.~Davies, and
  M.~Satyanarayanan, ``The impact of mobile multimedia applications on data
  center consolidation,'' in {\em Cloud Engineering (IC2E), 2013 IEEE
  International Conference on}, pp.~166--176, 2013.

\bibitem{mcmahan2011follow}
H.~B. McMahan, ``Follow-the-regularized-leader and mirror descent: Equivalence
  theorems and l1 regularization,'' 2011.

\end{thebibliography}
\bibliographystyle{ieeetr}
\begin{IEEEbiography}[{\includegraphics[width=1in,height=1.25in,clip,keepaspectratio]{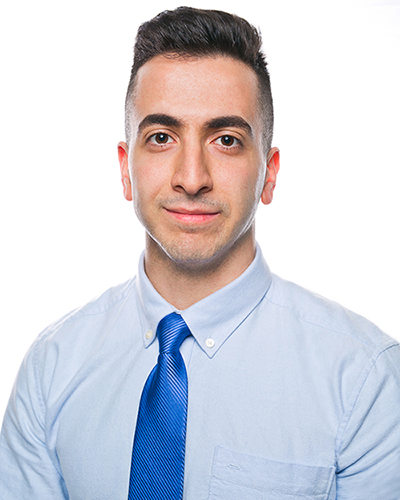}}]{Ashkan Yousefpour} (S'12-GS'13) is a Visiting Researcher at the University of California, Berkeley and a Ph.D. candidate at the Erik Jonsson School of Engineering and Computer Science, the University of Texas at Dallas. His research interests are in the areas of fog computing, cloud computing, Deep Learning, Internet of Things, and security. He obtained his B.S. in Computer Engineering from Amirkabir University of Technology, Iran, in 2013, and his M.S. in Computer Science from the University of Texas at Dallas, USA, in 2016. 
\end{IEEEbiography}
\begin{IEEEbiography}[{\includegraphics[width=1in,height=1.25in,clip,keepaspectratio]{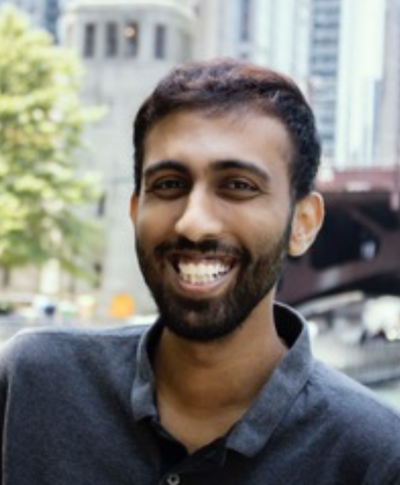}}]{Ashish~Patil} is a Master's student at the Erik Jonsson School of Engineering and Computer Science at the University of Texas at Dallas. His research interests are in the areas of Computer Networks, Fog Computing, Cloud Computing, and Security. He obtained his B.E. in Information Science from PES Institute of Technology, India, in 2017. \end{IEEEbiography}
\begin{IEEEbiography}[{\includegraphics[width=1in,height=1.25in,clip,keepaspectratio]{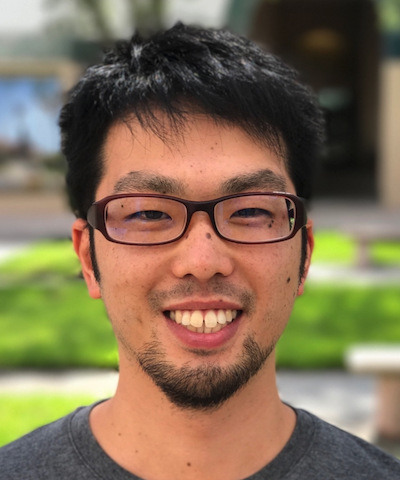}}]{Genya Ishigaki} (GS'14) received the B.S. and M.S. degrees in engineering from Soka University, Tokyo, Japan, in 2014 and 2016, respectively. He is currently pursuing the Ph.D. degree in computer science at Advanced Networks Research Laboratory, The University of Texas at Dallas, Richardson, TX, USA. His current research interests include design and recovery problems of interdependent networks, and software defined networking.
\end{IEEEbiography}
\begin{IEEEbiography}[{\includegraphics[width=1in,height=1.25in,clip,keepaspectratio]{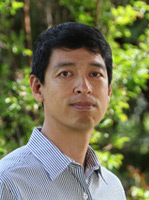}}]{Inwoong~Kim} has been a member of the research staff at Fujitsu Laboratories of America since 2007. His research focuses on optical networks and communications. He studied photonics and optical communications at the college of Optics \& Photonics, University of Central Florida, and received his Ph. D. degree in Optical Engineering in 2006. He worked as a postdoctoral fellow at CREOL. He received his M.S. degree in physics from KAIST in 1993. He worked in Korea Institute of Machinery and Materials from 1993 to 1998 and joined Ultra-short Pulse Laser Lab, KAIST, as a researcher in 1998.
\end{IEEEbiography}
\begin{IEEEbiography}[{\includegraphics[width=1in,height=1.25in,clip,keepaspectratio]{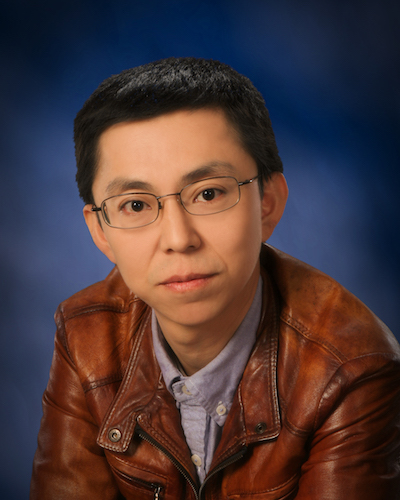}}]{Xi~Wang} is a Senior Researcher with Fujitsu Laboratories of America, where his research focuses on architecture design, control and management of Software Defined Networking, packet optical networks, vehicular networks, network programming, optical/wireless network integration, and future ICT fusion. Prior to joining Fujitsu, he was a visiting postdoctoral research associate with the University of Illinois at Chicago from 2005 to 2007, where he was engaged in research on innovative photonic networking technologies contributing to the design of a global optical infrastructure for eScience applications. He received B.S. degree in Electronic Engineering from Doshisha University, Japan in 1998, and M.E. and Ph.D. degrees in Information and Communication Engineering from the University of Tokyo, Japan in 2000 and 2003, respectively. 
\end{IEEEbiography}
\begin{IEEEbiography}[{\includegraphics[width=1in,height=1.25in,clip,keepaspectratio]{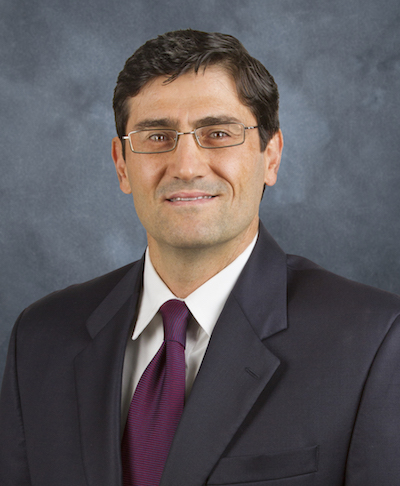}}]{Hakki~C.~Cankaya} is currently a solutions architect at Fujitsu Network Communications Inc. He is responsible for developing advanced networking solutions for a wide variety of Fujitsu customers. Prior to joining Fujitsu, Hakki worked in various senior research engineering positions at Bell Laboratories and Alcatel-Lucent, and was a member of the Alcatel-Lucent Technical Academy for several years. He is also an adjunct professor of computer science and electrical engineering at Southern Methodist University, Dallas, Texas.
\end{IEEEbiography}
\begin{IEEEbiography}[{\includegraphics[width=1in,height=1.25in,clip,keepaspectratio]{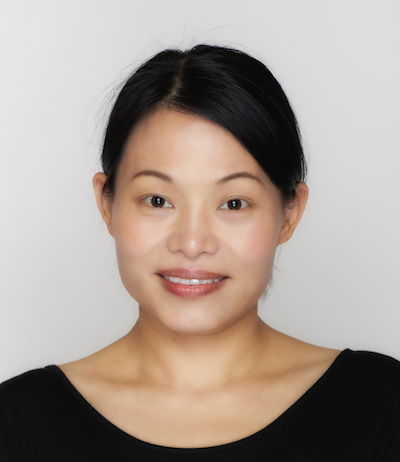}}]{Qiong~Zhang} received the Ph.D. and M.S. degrees from the University of Texas at Dallas in 2005 and 2000, respectively, and her B.S. in 1998 from Hunan University, all in Computer Science. She joined Fujitsu Laboratories of America in 2008. Before joining Fujitsu, she was an Assistant Professor in the department of Mathematical Sciences and Applied Computing at Arizona State University. Her research interests include optical networks, network design and modeling, network control and management, and distributed/cloud computing. She is the co-author of papers that received the Best Paper Award at IEEE GLOBECOM 2005, ONDM 2010, IEEE ICC 2011, and IEEE NetSoft 2016.  
\end{IEEEbiography}
\begin{IEEEbiography}[{\includegraphics[width=1in,height=1.25in,clip,keepaspectratio]{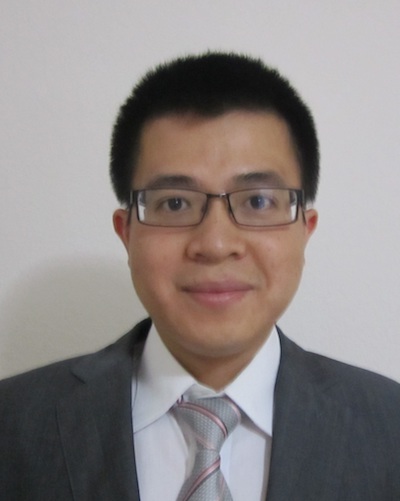}}]{Weisheng Xie} received the B.Eng. degree in Information Engineering from Beijing University of Posts and Telecommunications in 2010, the M.S. and Ph.D. degree in Telecommunications Engineering from the University of Texas at Dallas in 2013 and 2014, respectively. He is currently a Product Planner with Fujitsu Network Communications Inc., Richardson, TX. His research interests include optical network design and modeling, Software Defined Networking, and network virtualization.
\end{IEEEbiography}
\begin{IEEEbiography}[{\includegraphics[width=1in,height=1.25in,clip,keepaspectratio]{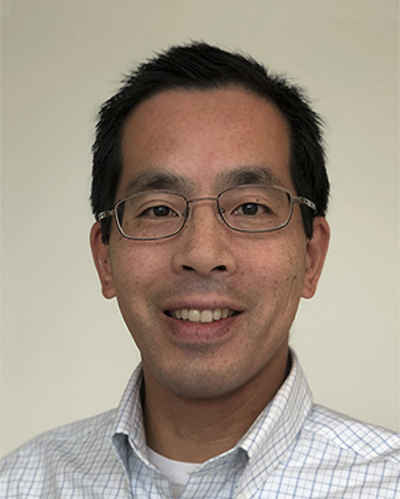}}]{Jason~P.~Jue} (M'99-SM'04) received the B.S. degree in Electrical Engineering and Computer Science from the University of California, Berkeley in 1990, the M.S. degree in Electrical Engineering from the University of California, Los Angeles in 1991, and the Ph.D. degree in Computer Engineering from the University of California, Davis in 1999. He is currently a Professor in the Department of Computer Science at the University of Texas at Dallas. His current research interests include optical networks and network survivability.
\end{IEEEbiography}
\end{document}